\documentclass[amsmath,amssymb,aps,pra,amsfonts,superscriptaddress,showkeys,showpacs,conference,12pt]{revtex4}
\usepackage{amsmath}
\usepackage{graphicx,amsmath}
\usepackage{booktabs}
\usepackage{amssymb}
\usepackage{graphicx}
\usepackage{subfigure}
\usepackage{phonetic}
\usepackage{extarrows}
\usepackage{float}
\usepackage{amsfonts}
\usepackage{epsfig}
\usepackage{bm}
\usepackage{mathrsfs}
\usepackage{bbm}
\usepackage{dsfont}
\usepackage{color}
\usepackage[colorlinks,citecolor=blue,linkcolor=blue]{hyperref}
\hypersetup{urlcolor=blue}

\date{\today}

\begin{document}
\title{Amplitude-modulation-based atom-mirror entanglement and mechanical squeezing in a hybrid optomechanical system}
\author{Cheng-Hua Bai}
\affiliation{Department of Physics, Harbin Institute of Technology, Harbin, Heilongjiang 150001, China}
\author{Dong-Yang Wang}
\affiliation{Department of Physics, Harbin Institute of Technology, Harbin, Heilongjiang 150001, China}
\author{Shou Zhang}
\email{szhang@ybu.edu.cn}
\affiliation{Department of Physics, Harbin Institute of Technology, Harbin, Heilongjiang 150001, China}
\affiliation{Department of Physics, College of Science, Yanbian University, Yanji, Jilin 133002, China}
\author{Shutian Liu}
\email{stliu@hit.edu.cn}
\affiliation{Department of Physics, Harbin Institute of Technology, Harbin, Heilongjiang 150001, China}
\author{Hong-Fu Wang}
\email{hfwang@ybu.edu.cn}
\affiliation{Department of Physics, College of Science, Yanbian University, Yanji, Jilin 133002, China}

\begin{abstract}
We consider a hybrid optomechanical system which is composed of the atomic ensemble and a standard optomechanical cavity driven by the periodically modulated external laser field. We investigate the asymptotic behaviors of Heisenberg operator first moments and clearly show the approaching process between the exact numerical results and analytical solutions. Based on the specific modulation forms of external driving and effective optomechanical coupling, we discuss in detail the atom-mirror entanglement enhancement, respectively. Compared with the constant driving regime, the entanglement can be greatly enhanced with more loose cavity decay rate and is more resistant to the thermal fluctuations of the mechanical bath. The desired form of periodically modulated effective optomechanical coupling can be precisely engineered by the external driving modulation components which can be derived analytically via Laplace transform. Meanwhile, resorting to the quantum interference mechanism caused by atomic ensemble and modulating the external driving  appropriately, the mechanical squeezing induced by the periodic modulation can be generated successfully in the unresolved regime.

\end{abstract}
\maketitle

\section{Introduction}\label{Sec1}
Recently, cavity optomechanics, as a controllable radiation-pressure interaction interface between cavity field and mechanical motion, has been raised more and more attention and research interests in both experimental and theoretical aspects~\cite{2007OE15017172,2009Physics,2013AnnPhys525215,2013CPB22114213,2014RMP861391,2018CPB27024204}. Particularly, with the fast-developing fields of microfabrication and nanotechnology, the significant progress of the experiments about cavity optomechanics has been made. So far, a variety of experimental structures can realize this controllable radiation-pressure interaction: whispering gallery microdisk and microsphere~\cite{2009Nature462633,2009NaturPhysics5489}, dielectric membrance~\cite{2008Nature45272}, nanorod~\cite{2009OE1712813}, silicon photonic waveguide~\cite{2008Nature456480}, and so forth. It is well known that the main potential applications about this subject are to test fundamentals of quantum mechanics for macroscopic systems~\cite{2013RMP85471} and to build quantum sensors for ultrahigh-precision measurements~\cite{2012PRA86053806,2014PRA90043825,2015PRA91063827,2015APL106121905,2017PRA95023844}. The generation of quantum entanglement between macroscopic objects has been a significantly important goal both in fundamental studies~\cite{2002JPCM14,2003RMP75715} and in numerous potential applications related to quantum computation, quantum communication, quantum information processing, etc~\cite{2003PRL9013,2005RMP77513,2011PRA84052327}. To this end, besides the entanglement generation between cavity field and mechanical oscillator~\cite{2007PRL98030405,2008PRA78032316,2011PRA84042342,2012PRA86042306}, many schemes have been put forward to generate the entanglement between two macroscopic oscillators~\cite{2009NJP11103044,2006PRL97150403,2011PRL107123601,2013PRA87022318,2014PRA89014302,2015NJP17103037,2016PRA94053807,2017PRA95043819,2017SR72545}. We also note that the remote quantum entanglement between two mechanical oscillators across two chips that are separated by 20 centimeters has been reported very recently~\cite{2018Nature556473}.

On the other hand, achieving mechanical squeezed state is also a task of paramount importance since it contributes to the realization of ultrahigh precision detection at or even below the standard quantum limit. In recent years, based on the cavity optomechanics, different methods to generate mechanical squeezing have been proposed, including quantum feedback control~\cite{2009PRA79052102}, parametric amplification and weak measurement~\cite{2011PRL107213603}, dissipative optomechanical coupling~\cite{2013PRA88013835}, quadratic optomechanical coupling~\cite{2014PRA89023849}, squeezing transfer from photons to phonons~\cite{2016PRA93043844}, nonlinearity~\cite{2015PRA91013834,2016SR624421,2016SR638559}, etc. In addition, the quantum squeezing of mechanical oscillator has also been observed in experiment~\cite{2015Science349952}.

In parallel with the development in cavity optomechanics, one modulation approach of particular interest is the so-called periodic modulation. The novel phenomena generated by applying periodic modulation to the cavity optomechanics have also been reported~\cite{2012NJP14075014,2013OE21020423,2014PRA89023843,2018PRA97042314,2018OE26013783,2018arxiv,2009PRL103213603,2012PRA86013820,2011PRA83033820,2018OE26011915,2018PRA97022336}. By appropriately modulating the time-periodic driving, Mari and Eisert have demonstrated that the entanglement between cavity field and mechanical oscillator can be greatly enhanced and large degrees of mechanical squeezing emerges in the resolved sideband regime~\cite{2009PRL103213603}. By further investigating and analyzing the interplay between the mechanical frequency modulation and input laser intensity modulation, it finds that an interference pattern presents and different choices of the relative phase between two modulations can either enhance or suppress the desired quantum effects in optomechanical system~\cite{2012PRA86013820}. Liao and Law proposed a method of reaching parametric resonance via modulating the driving field amplitude to generate the quadrature squeezing of mirrors~\cite{2011PRA83033820}. The method of simultaneously modulating the radiation-pressure coupling and mechanical spring constant to generate the ponderomotive squeezing and mechanical squeezing in the resolved sideband regime is also proposed recently~\cite{2018OE26011915}. We also note that by combing the periodic modulations for both the driving laser and the mechanical coupling strength simultaneously, entanglement dynamics of two coupled mechanical oscillators is investigated~\cite{2018PRA97022336}.

In this paper, we extend the optomechanical model in Ref.~\cite{2008PRA77050307} to the scenario of periodic modulation. We focus here on enhancing the atom-mirror entanglement for the specific modulation forms of external driving and effective optomechanical coupling, respectively. Compared to the constant driving regime, the periodic modulation greatly enhances the atom-mirror entanglement and its resistance to the thermal fluctuations of the mechanical bath with more loose cavity decay rate. In addition, the desired periodically modulated form of effective optomechanical coupling can be precisely engineered by the external driving modulation components which can be derived analytically via Laplace transform. Meanwhile, besides the entanglement illustrated in Ref.~\cite{2008PRA77050307}, the periodic modulation can also induce mechanical squeezing by appropriately modulating the external driving.  Resorting to the quantum interference mechanism caused by atomic ensemble, we generate the mechanical squeezing successfully in the unsolved sideband regime without the need of any feedback or additional squeezed light driving. We depict the Wigner functions at some different specific times and the time evolution of squeezing parameter and find that the mechanical oscillator is always squeezed but, due to the external driving modulation, the direction of squeezing rotates continuously in the phase space with the same period of modulation.

The rest of this paper is organized as follows. In Sec.~\ref{Sec2}, we describe the physical model and obtain the linearized dynamics of the system. In Sec.~\ref{Sec3}, we solve the dynamics of Heisenberg operator first moments analytically with a general periodically modulated amplitude in a perturbative way and derive the equation of motion for the correlation matrix which can completely describe the dynamics of the quantum fluctuations. In Sec.~\ref{Sec4}, we analyze the asymptotic behaviors of the first moments for the specific modulation form of external driving. We discuss in detail the atom-mirror entanglement enhancement based on the specific modulation forms of external driving and effective optomechanical coupling, respectively, in Sec.~\ref{Sec5}. Via appropriately modulating the external driving, we illustrate the generation of mechanical squeezing in the unresolved sideband regime in Sec.~\ref{Sec6}. Finally, we present our conclusions in Sec.~\ref{Sec7}.

\section{Model and Hamiltonian}\label{Sec2}

\begin{figure}[H]
\centering
\includegraphics[scale=1.0]{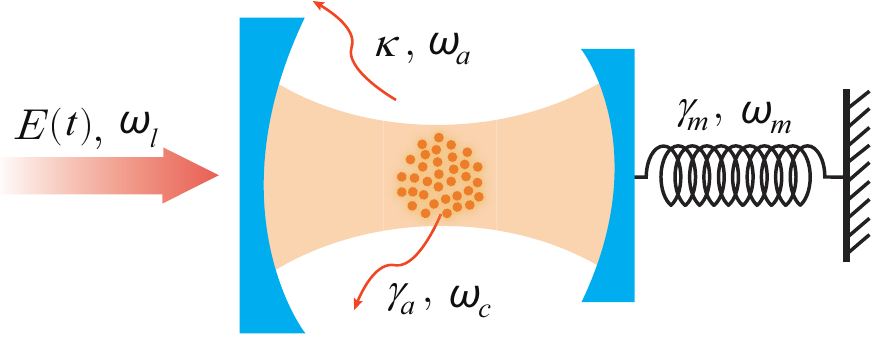}
\caption{(Color online) Schematic diagram of the considered hybrid optomechanical system. A cloud of identical two-level atoms is trapped in a standard optomechanical system, which is driven by an external laser field with periodically modulated amplitude $E(t)$.}\label{Fig1}
\end{figure}

The hybrid optomechanical system under consideration is depicted in Fig.~\ref{Fig1}, in which a cloud of identical two-level atoms (with frequency $\omega_c$ and decay rake $\gamma_a$) is trapped in a standard optomechanical system. An external laser filed with a time-dependent amplitude $E(t)$ and frequency $\omega_l$ drives the optomechanical cavity and the movable mirror modeled as mechanical oscillator with frequency $\omega_m$ and decay rate $\gamma_m$ is coupled to the optical field via the radiation-pressure interaction. The total Hamiltonian of the system (in the unit of $\hbar=1$) is given by the sum of a free evolution term
\begin{eqnarray}\label{E01}
H_0=\omega_aa^{\dag}a+\omega_cS_z+\frac{\omega_m}{2}(q^2+p^2),
\end{eqnarray}
and the interaction term
\begin{eqnarray}\label{E02}
H_{\mathrm{I}}=g_0(S_+a+S_-a^{\dag})-ga^{\dag}aq+iE(t)(e^{-i\omega_lt}a^{\dag}-e^{i\omega_lt}a).
\end{eqnarray}
Here, $a$ ($a^{\dag}$) is the annihilation (creation) operator of the cavity field (with frequency $\omega_a$ and decay rate $\kappa$); the collective spin operators of atoms are defined in terms of Pauli Matrices $S_{+,-,z}=\sum\limits_i\sigma_{+,-,z}^{(i)}$; and $q$ ($p$) is the dimensionless position (momentum) operator of the mechanical oscillator, satisfying the standard canonical commutation relation $[q, p]=i$. $g_0$ and $g$ refer to the atom-cavity coupling strength and the single-photon radiation-pressure coupling strength, respectively.

Under the conditions of sufficiently large atom number $N$ and weak atom-cavity coupling, the dynamics of the atomic polarization can be described in terms of a collective bosonic operator $c$ ($c^{\dag}$) in the Holstein-Primakoff representation~\cite{2008PRA77050307,2015PRA92033841}
\begin{eqnarray}\label{E03}
S_+=c^{\dag}\sqrt{N-c^{\dag}c}\simeq\sqrt{N}c^{\dag},~~~~~S_-=\sqrt{N-c^{\dag}c}c\simeq\sqrt{N}c,~~~~~S_z=c^{\dag}c-N/2.
\end{eqnarray}
In the rotating frame with respect to laser frequency $\omega_l$, the total Hamiltonian is rewritten as follows:
\begin{eqnarray}\label{E04}
H=\delta_aa^{\dag}a+\Delta_cc^{\dag}c+\frac{\omega_m}{2}(q^2+p^2)+G_0(c^{\dag}a+ca^{\dag})-ga^{\dag}aq+iE(t)(a^{\dag}-a),
\end{eqnarray}
where $\delta_a=\omega_a-\omega_l$ and $\Delta_c=\omega_c-\omega_l$ are, respectively, the cavity and atomic detuning with respect to the laser. $G_0=\sqrt{N}g_0$ is the collective atom-cavity coupling strength. The time-dependent amplitude $E(t)$ is imposed the structure of a periodic modulation such that $E(t+\tau)=E(t)$ for some $\tau>0$ of the order of $\omega_m^{-1}$, though the particular form is left unspecified yet.

In addition to the coherent dynamics, the system is also unavoidably affected by the fluctuation-dissipation processes resulting from the environment. Taking all the damping and noise terms into account, the dynamics of the system is completely described by the following set of nonlinear quantum Langevin equations (QLEs)~\cite{Book1}:
\begin{eqnarray}\label{E05}
\frac{dq}{dt}&=&\omega_mp, \cr\cr
\frac{dp}{dt}&=&-\omega_mq-\gamma_mp+ga^{\dag}a+\xi(t), \cr\cr
\frac{da}{dt}&=&-(\kappa+i\delta_a)a+igaq-iG_0c+E(t)+\sqrt{2\kappa}a_{\mathrm{in}}(t), \cr\cr
\frac{dc}{dt}&=&-(\gamma_a+i\Delta_c)c-iG_0a+\sqrt{2\gamma_a}F_c(t),
\end{eqnarray}
where $a_{\mathrm{in}}(t)$ and $F_c(t)$ are the zero-mean noise operators for cavity and atoms, respectively, with the only nonzero correlation functions
\begin{eqnarray}\label{E06}
\langle a_{\mathrm{in}}(t)a_{\mathrm{in}}^{\dag}(t^{\prime})\rangle&=&\delta(t-t^{\prime}), \cr\cr
\langle F_c(t)F_c^{\dag}(t^{\prime})\rangle&=&\delta(t-t^{\prime}).
\end{eqnarray}
$\xi(t)$ is the stochastic Hermitian Brownian noise operator describing the dissipative friction forces subjecting to the mechanical oscillator and its non-Markovian correlation function is given by~\cite{2001PRA63023812}
\begin{eqnarray}\label{E07}
\langle\xi(t)\xi(t^{\prime})\rangle=\frac{\gamma_m}{2\pi\omega_m}\int\omega\left[\coth\left(\frac{\hbar\omega}{2k_BT}\right)+1\right]e^{-i\omega(t-t^{\prime})}d\omega,
\end{eqnarray}
where $k_B$ is the Boltzmann constant and $T$ is the temperature of the mechanical bath. However, for a high quality mechanical oscillator with $Q=\omega_m/\gamma_m\gg1$, the colored spectrum of Eq.~(\ref{E07}) acquires the Markovian character~\cite{Note1}
\begin{eqnarray}\label{E08}
\langle\{\xi(t), \xi(t^{\prime})\}\rangle\simeq2\gamma_m\coth\left(\frac{\hbar\omega_m}{2k_BT}\right)\delta(t-t^{\prime})
=2\gamma_m(2n_{\mathrm{th}}+1)\delta(t-t^{\prime}),
\end{eqnarray}
where $n_{\mathrm{th}}=[\mathrm{exp}(\hbar\omega_m/k_BT)-1]^{-1}$ is the mean occupation number of the mechanical mode and $\{\cdots, \cdots\}$ is the anticommutator.

In general, the set of coupled nonlinear QLEs in Eq.~(\ref{E05}) is difficult to solve directly. However, when the system is strongly driven to a large first moment by the external laser, one can rewrite each Heisenberg operator as follows: $\mathscr{O}(t)=\langle\mathscr{O}(t)\rangle+\delta\mathscr{O}(t)$ ($\mathscr{O}=q, p, a, c$), where $\delta\mathscr{O}(t)$ are quantum fluctuation operators with zero-mean around the classical $c$-number first moments $\langle\mathscr{O}(t)\rangle$. Thus, the standard linearization techniques can be applied to the nonlinear QLEs in Eq.~(\ref{E05}).  The equation of motion for Heisenberg operator first moments of system is given by the following set of nonlinear differential equations:
\begin{eqnarray}\label{E09}
\langle\dot{q}(t)\rangle&=&\omega_m\langle p(t)\rangle, \cr\cr
\langle\dot{p}(t)\rangle&=&-\omega_m\langle q(t)\rangle-\gamma_m\langle p(t)\rangle+g|\langle a(t)\rangle|^2, \cr\cr
\langle\dot{a}(t)\rangle&=&-[\kappa+i(\delta_a-g\langle q(t)\rangle)]\langle a(t)\rangle-iG_0\langle c(t)\rangle+E(t), \cr\cr
\langle\dot{c}(t)\rangle&=&-(\gamma_a+i\Delta_c)\langle c(t)\rangle-iG_0\langle a(t)\rangle.
\end{eqnarray}
The linearized QLEs for the quantum fluctuation operators are
\begin{eqnarray}\label{E10}
\frac{d\delta q}{dt}&=&\omega_m\delta p, \cr\cr
\frac{d\delta p}{dt}&=&-\omega_m\delta q-\gamma_m\delta p+g\langle a(t)\rangle^{\ast}\delta a+g\langle a(t)\rangle\delta a^{\dag}+\xi(t), \cr\cr
\frac{d\delta a}{dt}&=&-[\kappa+i(\delta_a-g\langle q(t)\rangle)]\delta a+ig\langle a(t)\rangle\delta q-iG_0\delta c+\sqrt{2\kappa}a_{\mathrm{in}}(t), \cr\cr
\frac{d\delta c}{dt}&=&-(\gamma_a+i\Delta_c)\delta c-iG_0\delta a+\sqrt{2\gamma_a}F_c(t),
\end{eqnarray}
and the corresponding linearized system Hamiltonian for the quantum fluctuation operators reads
\begin{eqnarray}\label{E11}
H&=&(\delta_a-g\langle q(t)\rangle)\delta a^{\dag}\delta a+\Delta_c\delta c^{\dag}\delta c+\frac{\omega_m}{2}(\delta q^2+\delta p^2) \cr\cr
&&+G_0(\delta c^{\dag}\delta a+\delta c\delta a^{\dag})-g(\langle a(t)\rangle^{\ast}\delta a+\langle a(t)\rangle\delta a^{\dag})\delta q.
\end{eqnarray}

\section{Dynamics of Heisenberg operator first moments and quantum fluctuations}\label{Sec3}
The classical evolution of the system is governed by the dynamics of the first moments in Eq.~(\ref{E09}). One can get the time evolution of the first moments by numerically solving the set of differential equations in Eq.~(\ref{E09}) although it is nonlinear. However, when the system is far away from the regions of optomechanical instabilities and multistabilities~\cite{2008NJP10095013}, the radiation pressure coupling can be treated in terms of perturbation. Besides, since $E(t+\tau)=E(t)$, according to Floquet's theory, the asymptotic solutions of first moments will have the same periodicity of modulation in the long time limit: $\langle\mathscr{O}(t+\tau)\rangle=\langle\mathscr{O}(t)\rangle$. Thus one can perform a double expansion of the asymptotic solutions $\langle\mathscr{O}(t)\rangle$ in powers of the radiation pressure coupling strength $g$ and in terms of Fourier components:
\begin{eqnarray}\label{E12}
\langle\mathscr{O}(t)\rangle=\sum_{j=0}^{\infty}\sum_{n=-\infty}^{\infty}\mathscr{O}_{n,j}e^{in\Omega t}g^j,
\end{eqnarray}
where $n$($j$) are integers and $\Omega=2\pi/\tau$ is the fundamental modulation frequency. In addition, the periodic driving amplitude $E(t)$ can also be expanded in terms of the similar Fourier series
\begin{eqnarray}\label{E13}
E(t)=\sum_{n=-\infty}^{\infty}E_ne^{-in\Omega t}.
\end{eqnarray}
Substituting Eqs.~(\ref{E12}) and (\ref{E13}) into Eq.~(\ref{E09}), the time-independent coefficients $\mathscr{O}_{n,j}$ in Eq.~(\ref{E12}) are obtained by the following set of recursive relations:
\begin{eqnarray}\label{E14}
q_{n,0}&=&0, ~~~~~~~~~~~~p_{n,0}=0, \cr\cr
a_{n,0}&=&\frac{[i(n\Omega+\Delta_c)+\gamma_a]E_{-n}}{[i(n\Omega+\delta_a)+\kappa]\cdot[i(n\Omega+\Delta_c)+\gamma_a]+G_0^2}, \cr\cr
c_{n,0}&=&\frac{G_0E_{-n}}{i\{[i(n\Omega+\delta_a)+\kappa]\cdot[i(n\Omega+\Delta_c)+\gamma_a]+G_0^2\}},
\end{eqnarray}
corresponding to the zero-order perturbation with respect to $g$, and for all $j\gg1$,
\begin{eqnarray}\label{E15}
p_{n,j}&=&\frac{in\Omega}{\omega_m}q_{n,j}, \cr\cr
q_{n,j}&=&\omega_m\sum_{k=0}^{j-1}\sum_{m=-\infty}^{\infty}\frac{a_{m,k}^{\ast}a_{n+m,j-k-1}}{\omega_m^2-(n\Omega)^2+i\gamma_mn\Omega}, \cr\cr
a_{n,j}&=&i[\gamma_a+i(\Delta_c+n\Omega)]\sum_{k=0}^{j-1}\sum_{m=-\infty}^{\infty}\frac{a_{m,k}q_{n-m,j-k-1}}{[\kappa+i(\delta_a+n\Omega)]\cdot[\gamma_a+i(\Delta_c+n\Omega)]+G_0^2}, \cr\cr
c_{n,j}&=&G_0\sum_{k=0}^{j-1}\sum_{m=-\infty}^{\infty}\frac{a_{m,k}q_{n-m,j-k-1}}{[\kappa+i(\delta_a+n\Omega)]\cdot[\gamma_a+i(\Delta_c+n\Omega)]+G_0^2}.
\end{eqnarray}
For all the calculations carried out in this paper, we truncated the analytical solutions up to $j\leq6$ and $|n|\leq5$ so that the level of approximation is high enough to agree well with the exact numerical solutions.

On the other hand, as long as the time evolution of the first moments is obtained, the dynamics of the corresponding quantum fluctuations is easy to be solved. To this end, it is convenient to introduce the quadrature operators and corresponding Hermitian input noise operators for the cavity field and atoms, respectively,
\begin{eqnarray}\label{E16}
\delta X&=&(\delta a+\delta a^{\dag})/\sqrt{2},~~~~~
\delta Y=(\delta a-\delta a^{\dag})/\sqrt{2}i, \cr\cr
\delta x&=&(\delta c+\delta c^{\dag})/\sqrt{2},~~~~~
\delta y=(\delta c-\delta c^{\dag})/\sqrt{2}i, \cr\cr
X_{\mathrm{in}}&=&(a_{\mathrm{in}}+a_{\mathrm{in}}^{\dag})/\sqrt{2},~~~~~
Y_{\mathrm{in}}=(a_{\mathrm{in}}-a_{\mathrm{in}}^{\dag})/\sqrt{2}i, \cr\cr
x_{\mathrm{in}}&=&(F_c+F_c^{\dag})/\sqrt{2},~~~~~
y_{\mathrm{in}}=(F_c-F_c^{\dag})/\sqrt{2}i,
\end{eqnarray}
and the vectors of quadrature fluctuation operators and corresponding noises are
\begin{eqnarray}\label{E17}
u(t)&=&[\delta q, \delta p, \delta X, \delta Y, \delta x, \delta y]^T, \cr\cr
n(t)&=&[0, \xi(t), \sqrt{2\kappa}X_{\mathrm{in}}(t), \sqrt{2\kappa}Y_{\mathrm{in}}(t), \sqrt{2\gamma_a}x_{\mathrm{in}}(t), \sqrt{2\gamma_a}y_{\mathrm{in}}(t)]^T.
\end{eqnarray}
So the linearized QLEs which govern the dynamics of the quantum fluctuations can be written in a compact form:
\begin{eqnarray}\label{E18}
\frac{du}{dt}=A(t)u+n(t),
\end{eqnarray}
in which $A(t)$ is a $6\times6$ time-dependent matrix:
\begin{eqnarray}\label{E19}
A(t)=
\begin{bmatrix}
0 & ~\omega_m & ~0 & ~0 & ~0 & ~0 \\
-\omega_m & ~-\gamma_m & ~G_x(t) & ~G_y(t) & ~0 & ~0 \\
-G_y(t) & ~0 & ~-\kappa & ~\Delta_a(t) & ~0 & ~G_0 \\
G_x(t) & ~0 & ~-\Delta_a(t) & ~-\kappa & ~-G_0 & ~0 \\
0 & ~0 & ~0 & ~G_0 & ~-\gamma_a & ~\Delta_c \\
0 & ~0 & ~-G_0 & ~0 & ~-\Delta_c & ~-\gamma_a
\end{bmatrix},
\end{eqnarray}
where 
\begin{eqnarray}\label{E20}
\Delta_a(t)=\delta_a-g\langle q(t)\rangle
\end{eqnarray}
is the effective time-modulated detuning, and $G_x(t)$ and $G_y(t)$ are the real and imaginary parts of the effective optomechanical coupling, respectively, 
\begin{eqnarray}\label{E21}
G(t)=\sqrt{2}g\langle a(t)\rangle.
\end{eqnarray}

Thanks to the linearized dynamics for the quantum fluctuations and the zero-mean Gaussian nature for the quantum noises, the time evolution of the quantum fluctuations can be completely described by the $6\times6$ correlation matrix (CM) $V(t)$ whose matrix element is defined by
\begin{eqnarray}\label{E22}
V_{k,l}=\langle u_{k}(t)u_{l}(t)+u_{l}(t)u_{k}(t)\rangle/2.
\end{eqnarray}
The equation of motion for the CM can be derived from Eqs.~(\ref{E18}) and (\ref{E22}) (see Appendix \ref{App1})
\begin{eqnarray}\label{E23}
\frac{dV}{dt}=A(t)V(t)+V(t)A^T(t)+D,
\end{eqnarray}
where $A^T(t)$ refers to the transpose of $A(t)$, and $D$ is a diffusion matrix whose matrix element is related to the noise correlations and defined by
\begin{eqnarray}\label{E24}
\delta(t-t^{\prime})D_{k, l}=\langle n_k(t)n_l(t^{\prime})+n_l(t^{\prime})n_k(t)\rangle/2,
\end{eqnarray}
from the correlation functions of Eqs.~(\ref{E06}) and (\ref{E08}), the diagonal matrix $D$ is
\begin{eqnarray}\label{E25}
D=\mathrm{diag}[0, \gamma_m(2n_{\mathrm{th}}+1), \kappa, \kappa, \gamma_a, \gamma_a].
\end{eqnarray}

Until now, the dynamics of the quantum fluctuations in Eq.~(\ref{E10}) is completely described by the equation of motion for the CM in Eq.~(\ref{E23}).

\section{Asymptotic behaviors of Heisenberg operator first moments for the specific modulation form of external driving}\label{Sec4}
In the long time asymptotic regime, the time-dependent matrix $A(t)$ will have the same periodicity of the modulation, i.e., $A(t+\tau)=A(t)$. Moreover, the stability of the system requires that all of the eigenvalues of $A(t)$ have negative real parts for all time $t$. In what follows, the stability condition should be carefully guaranteed anytime.

\begin{figure}
\centering
\includegraphics[scale=0.5]{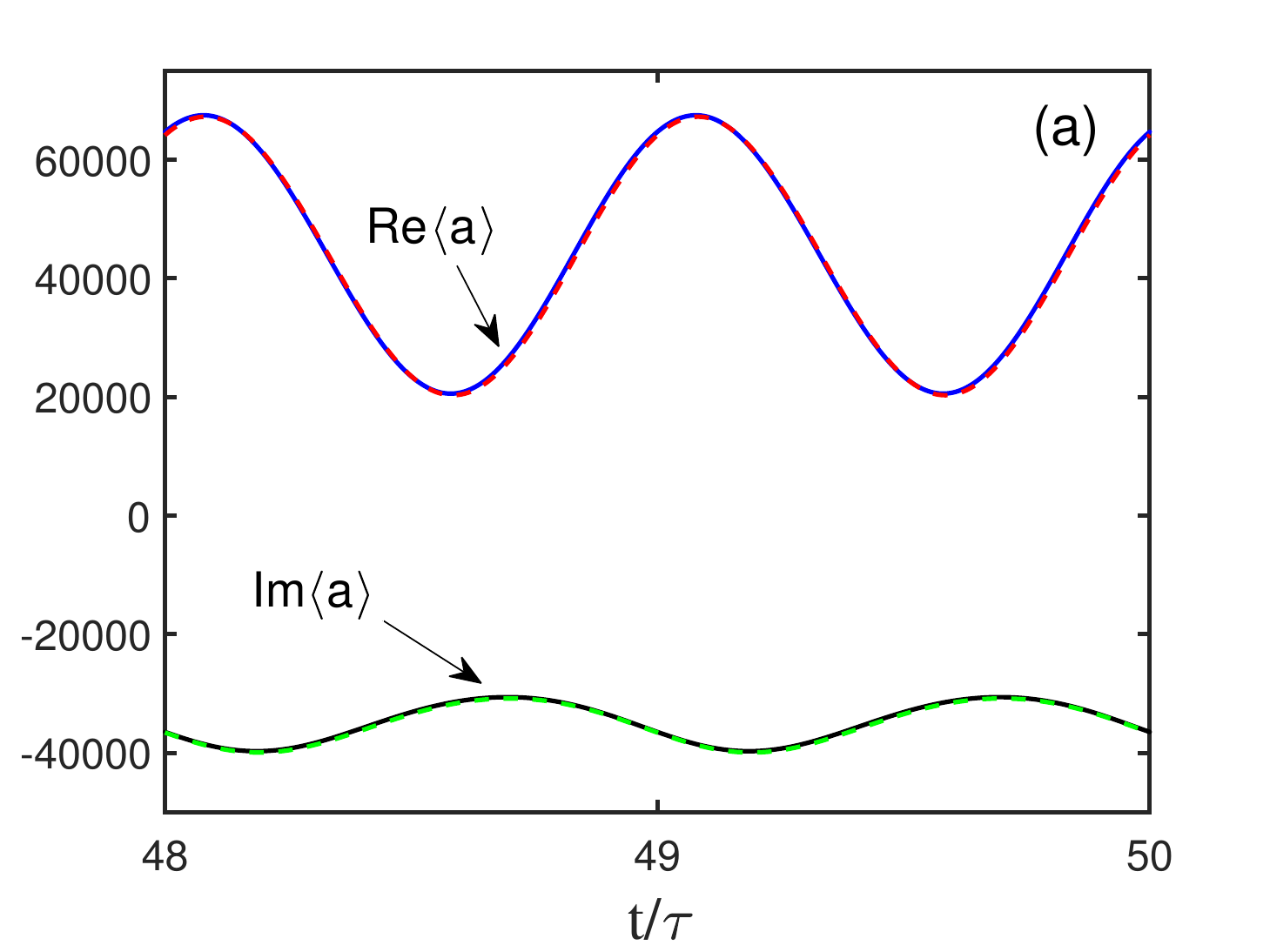}
\hspace{0.3in}
\includegraphics[scale=0.5]{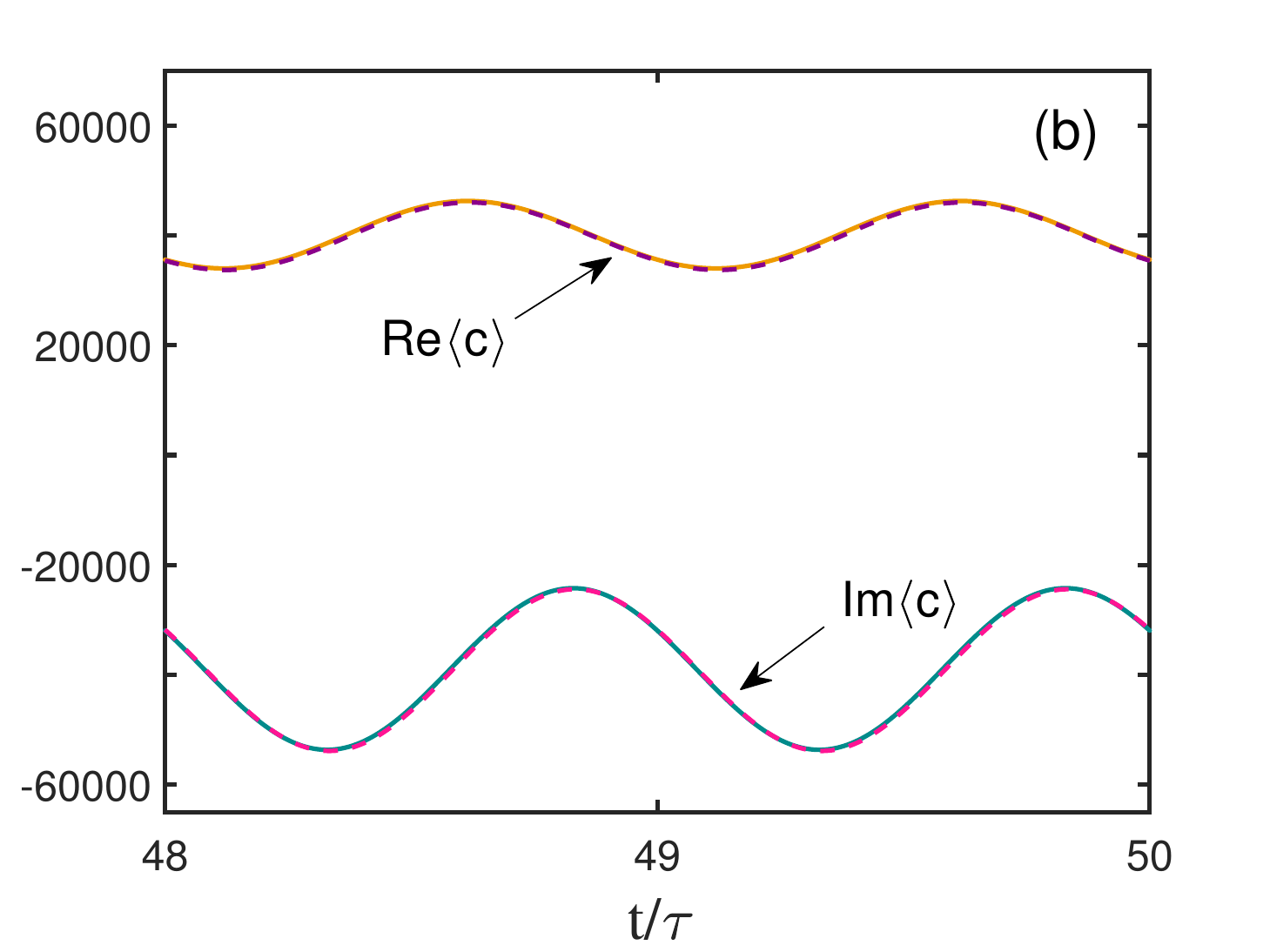}
\caption{(Color online) Time evolution of the real and imaginary parts of (a) the first moment $\langle a(t)\rangle$ and (b) the first moment $\langle c(t)\rangle$ for the time interval of two modulation periods $[48\tau, 50\tau]$. In both figures, the solid and dashed lines represent, respectively, the exact numerical results and analytical solutions. The chosen system parameters are (in units of $\omega_m$): $\delta_a=1$, $\kappa=2$, $\gamma_m=10^{-3}$, $g=10^{-5}$, $\Delta_c=-1$, $\gamma_a=0.1$, $G_0=1$, $E_0=15\times10^4$, $E_{1}=E_{-1}=3\times10^4$, and the modulation frequency $\Omega=2$.}\label{Fig2}
\end{figure}

\begin{figure}
\centering
\includegraphics[scale=0.5]{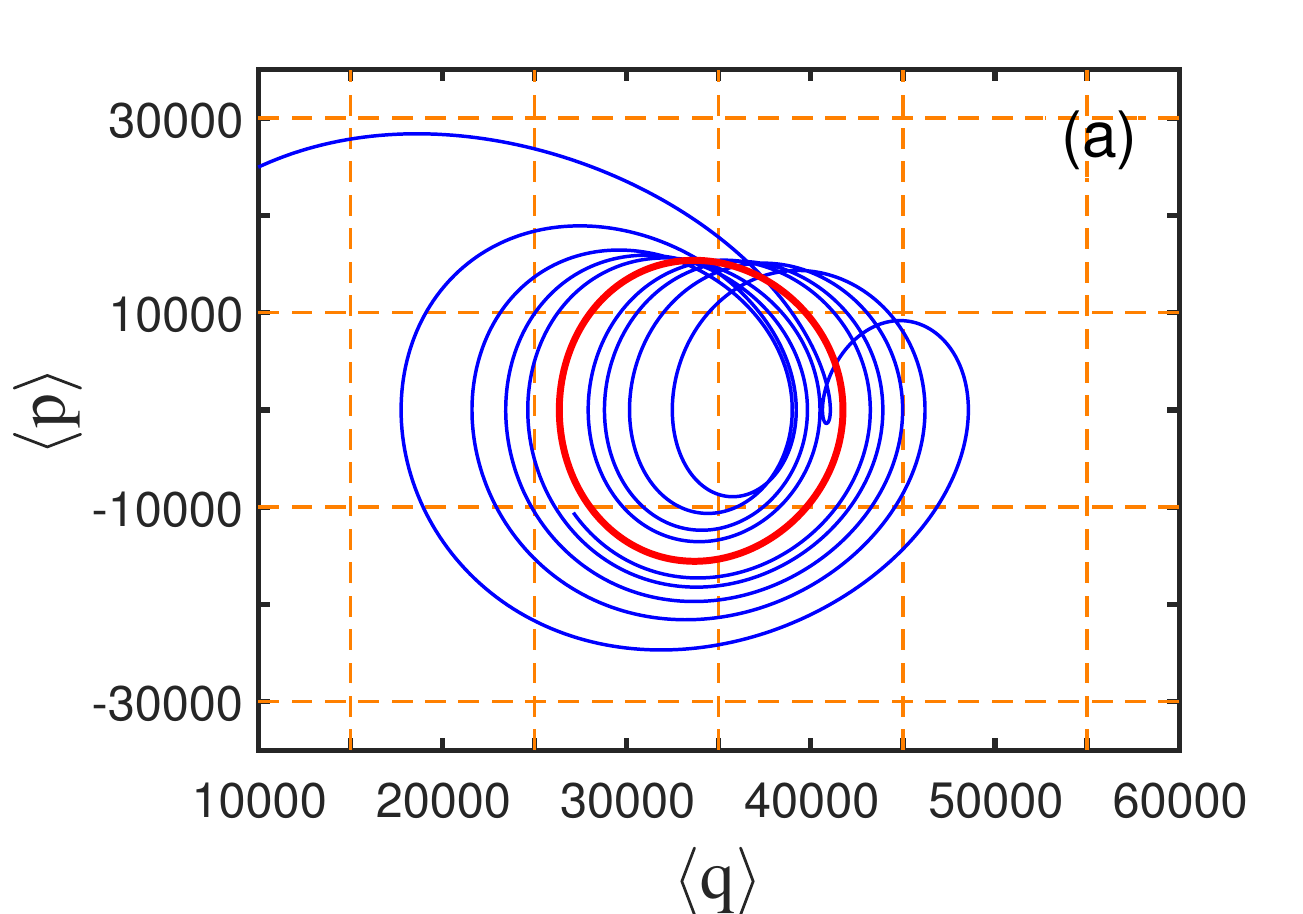}
\hspace{0.3in}
\includegraphics[scale=0.5]{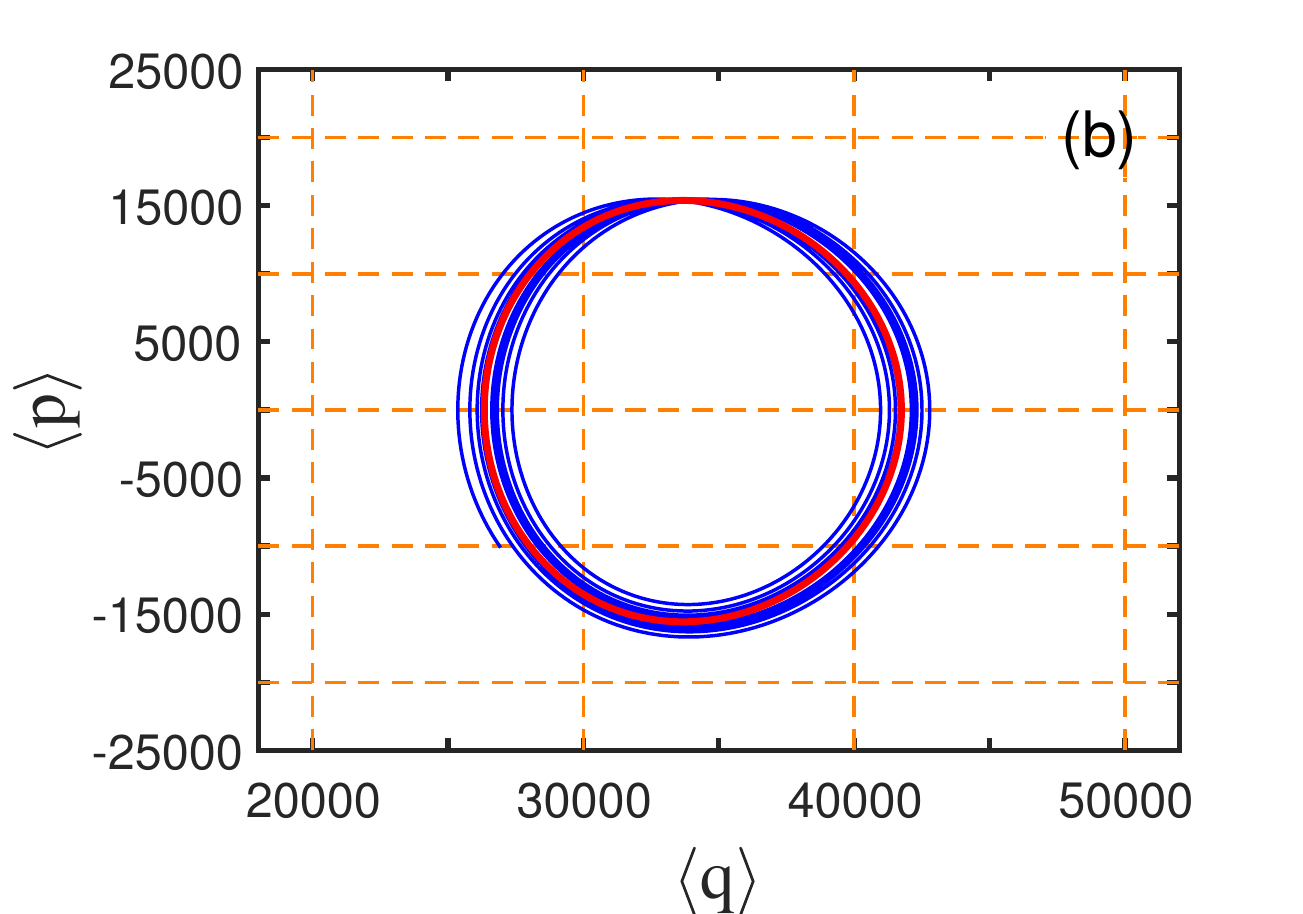}
\vspace{0.10in}
\includegraphics[scale=0.5]{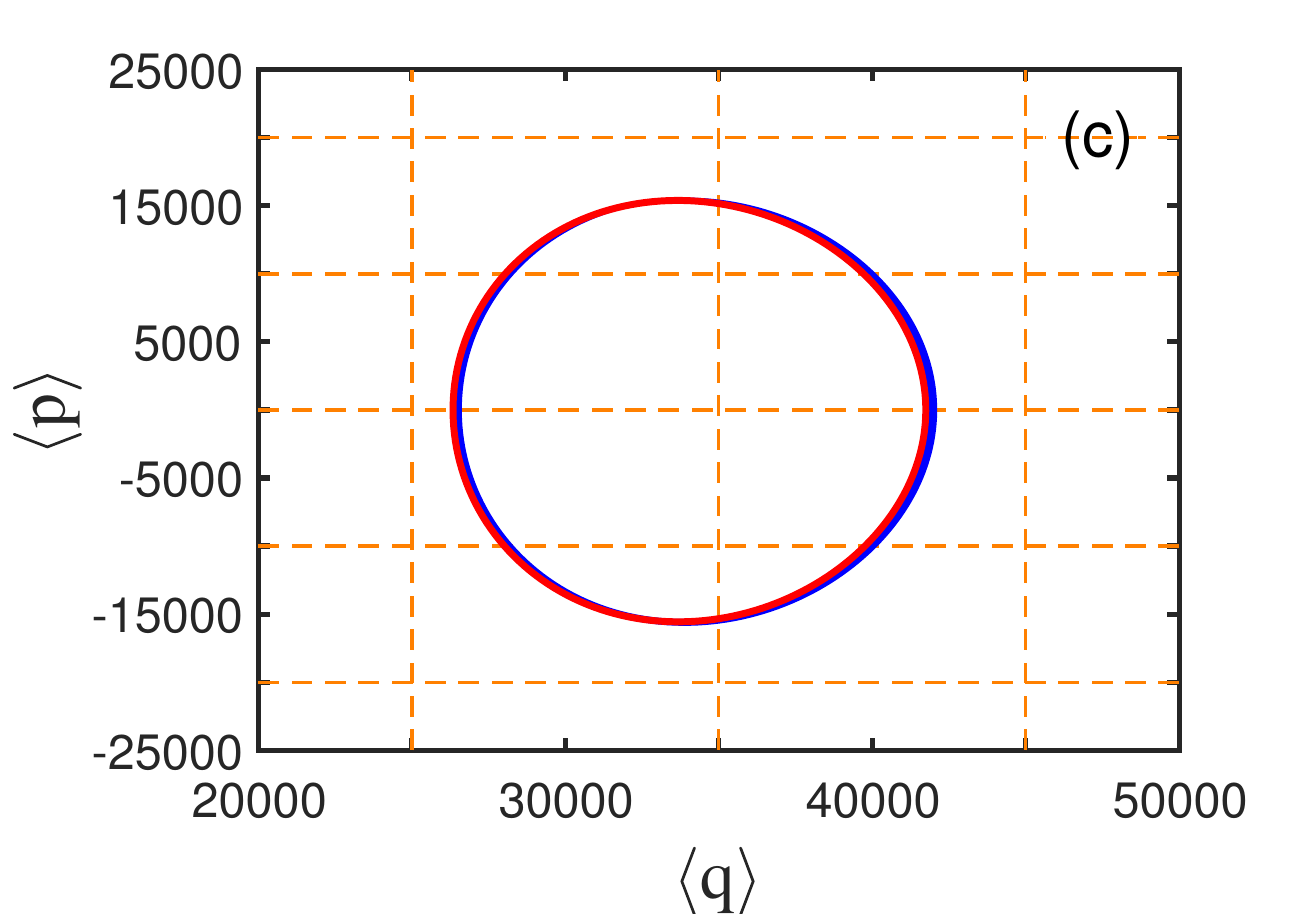}
\hspace{0.3in}
\includegraphics[scale=0.5]{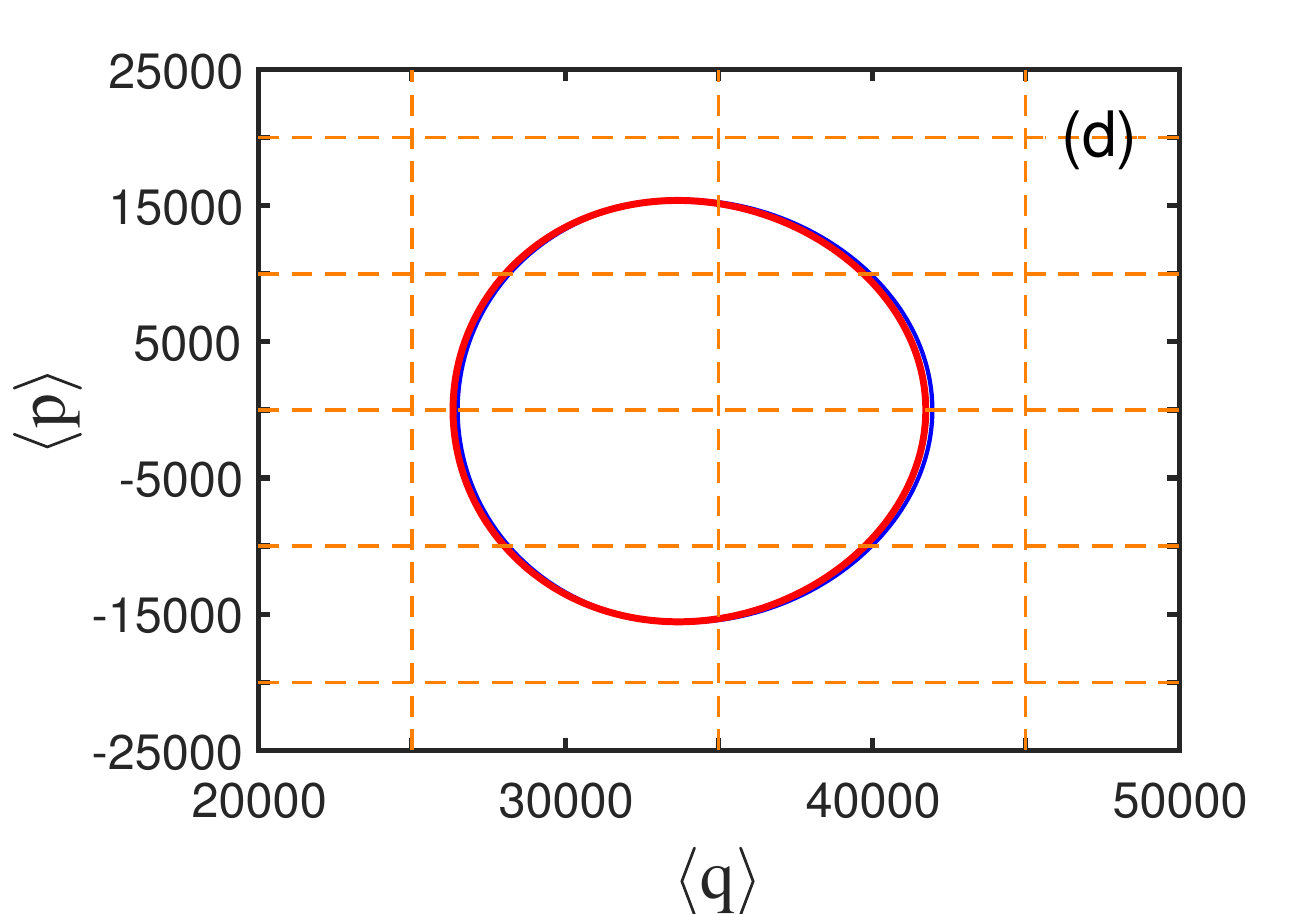}
\caption{(Color online) Phase space trajectories of the first moments for the mechanical oscillator $\langle q(t)\rangle$ and $\langle p(t)\rangle$ for time intervals (a) $[0, 10\tau]$, (b) $[10\tau, 20\tau]$, $[20\tau, 30\tau]$, and (d) $[30\tau, 40\tau]$. In all figures, the blue and red solid lines are obtained from the exact numerical results and analytical solutions, respectively. The system parameters are the same as those in Fig.~\ref{Fig2}.}\label{Fig3}
\end{figure}

To proceed, we investigate the asymptotic behaviors of Heisenberg operator first moments for the specific modulation form of external driving. As mentioned above, in the long time limit, the Heisenberg operator first moments $\langle\mathscr{O}(t)\rangle$ ($\mathscr{O}=q, p, a, c$) will have the same periodicity of the performed modulation. For simplicity, we restrict the modulation form of external driving $E(t)$ only to the lowest-amplitude components: $E(t)=E_{-1}e^{i\Omega t}+E_0+E_1e^{-i\Omega t}$. In Fig.~\ref{Fig2}, we plot the asymptotic time evolution of the real and imaginary parts of the first moments $\langle a(t)\rangle$ and $\langle c(t)\rangle$. One can clearly see that the first moments $\langle a(t)\rangle$ and $\langle c(t)\rangle$ are indeed $\tau$ periodicity in the long time limit. We also check the approximation validity of the analytical solutions in Eqs.~(\ref{E14}) and (\ref{E15}) via comparing with the exact numerical results obtained through Eq.~(\ref{E09}). We find that, after about 50 modulation periods, the analytical solutions agree with the exact numerical results very well.

Due to the asymptotic periodicity, the trajectories of the first moments in the phase space will converge to a limit cycle in the end. For this reason, to gain more insights about this characteristic and clearly show the asymptotic behaviors, as illustrated in Fig.~\ref{Fig3}, we plot the phase space trajectories of the first moments for the mechanical oscillator $\langle q(t)\rangle$ and $\langle p(t)\rangle$ in the particular time intervals. One can find that, with the periodic modulation proceeding, the trajectories indeed gradually converge to a limit cycle which is well approximated by the analytical prediction. Figure~\ref{Fig3} also clearly presents the slow asymptotic process between the exact numerical results and analytical solutions. The system thus obtains the same period of the performed modulation in the long time limit.

\section{Atom-mirror entanglement enhancement}\label{Sec5}
\subsection{Atom-mirror entanglement enhancement for the specific modulation form of external driving}
In this section, we investigate the atom-mirror entanglement enhancement under the specific modulation form of external driving. Since the asymptotic state of the system is Gaussian, it is very convenient to introduce the logarithmic negativity $E_N$ to measure the entanglement~\cite{2002PRA65032314,2004PRA70022318}, which can be readily computed from the reduced $4\times4$ CM $V_{\mathrm{am}}(t)$ for the collective atomic mode and mechanical mode. $V_{\mathrm{am}}(t)$ can be obtained from the full $6\times6$ CM $V(t)$ by just extracting the first two and last two rows and columns. If the reduced CM $V_{\mathrm{am}}(t)$ is written in the following form:
\begin{eqnarray}\label{E26}
V_{\mathrm{am}}(t)=
\begin{bmatrix}
A~ & ~C \\
C^T~ & ~B
\end{bmatrix},
\end{eqnarray}
where $A$, $B$, and $C$ are $2\times2$ subblock matrices of $V_{\mathrm{am}}(t)$, which describe the local properties of mechanical mode,  collective atomic mode, and the intermode correlation between them, respectively, then $E_N$ is defined as
\begin{eqnarray}\label{E27}
E_N=\mathrm{max}[0, -\ln(2\eta^-)],
\end{eqnarray}
with
\begin{eqnarray}\label{E28}
\eta^-&\equiv&2^{-1/2}\left\{\Sigma-\left[\Sigma^2-4\mathrm{det}V_{\mathrm{am}}\right]^{1/2}\right\}^{1/2}, \cr\cr
\Sigma&\equiv&\mathrm{det}A+\mathrm{det}B-2\mathrm{det}C.
\end{eqnarray}
The collective atomic mode and mechanical mode are said to be entangled ($E_N>0$) if and only if $\eta^-<1/2$ which is equivalent to Simon's necessary and sufficient nonpositive partial transpose criteria~\cite{2000PRL842726}.
 
\begin{figure}[H]
\centering
\includegraphics[scale=0.5]{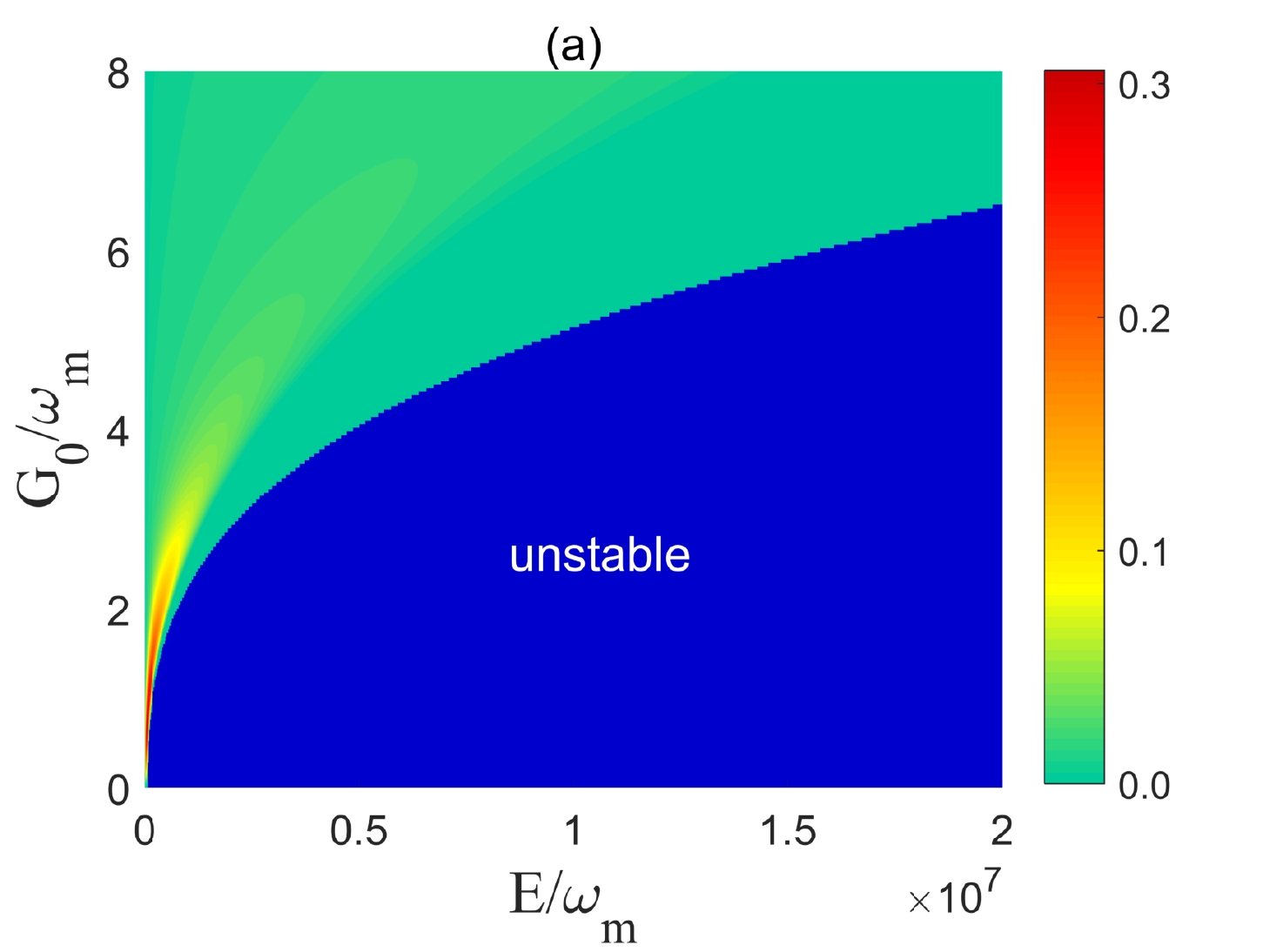}
\hspace{0.3in}
\includegraphics[scale=0.5]{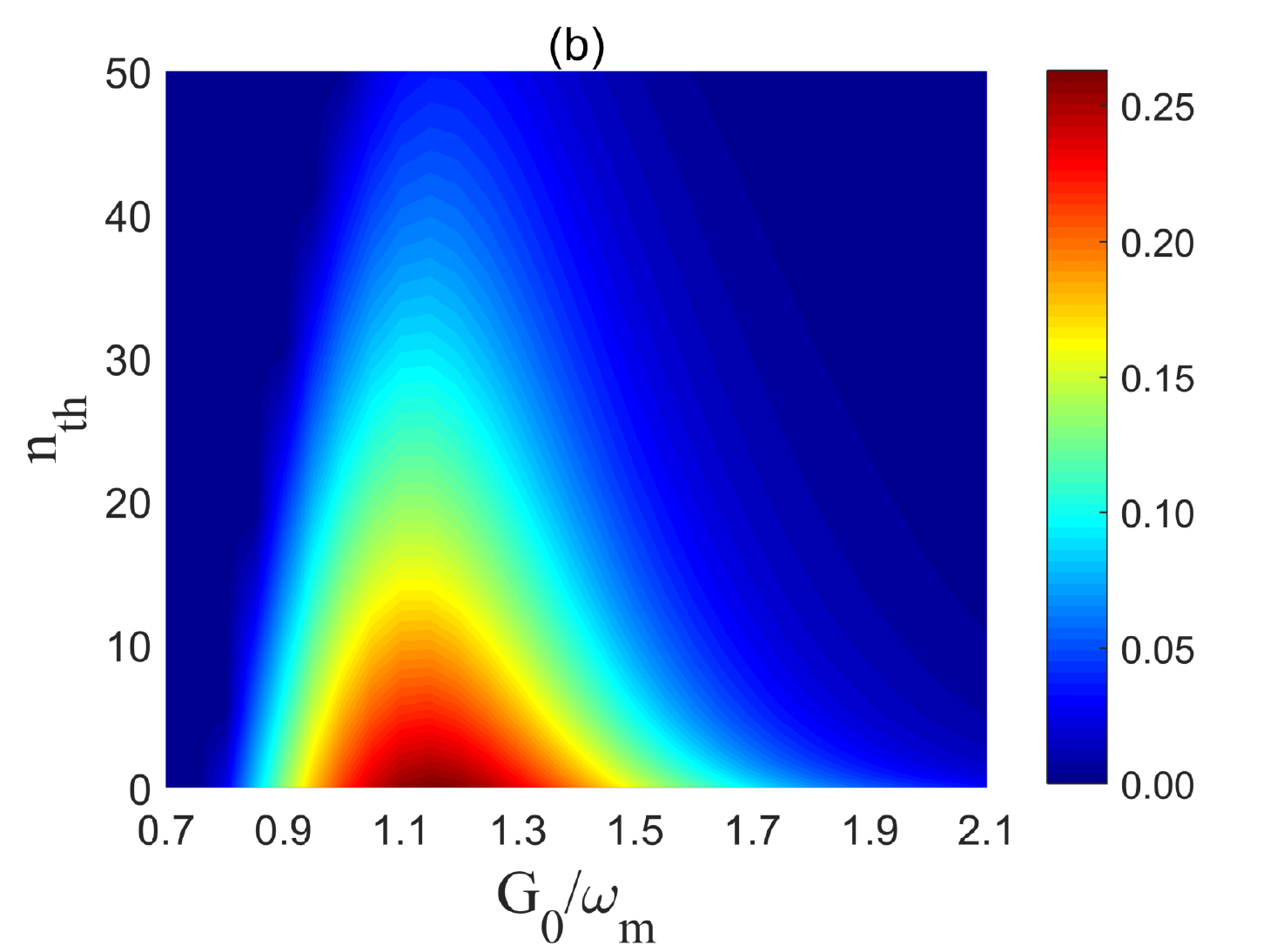}
\caption{(Color online) Steady-state atom-mirror entanglement $E_N$ without modulation $(\Omega=0)$. (a) $E_N$ versus the time-independent driving amplitude $E$ and atom-cavity coupling strength $G_0$ with the mean thermal phonon occupation number $n_{\mathrm{th}}=0$ and the effective cavity detuning $\Delta_a=\omega_m$. In the blue region, the system shows unstable behavior. (b) $E_N$ versus the atom-cavity coupling strength $G_0$ and mean thermal phonon occupation number $n_{\mathrm{th}}$ with $\Delta_a=\omega_m$ and $E=1.2\times10^5\omega_m$. In both figures, the cavity decay rate $\kappa=0.2\omega_m$ and other parameters are the same as those in Fig.~\ref{Fig2}.}\label{Fig4}	
\end{figure}

For comparison, we first present the steady-state entanglement behavior when there is no periodic modulation ($\Omega=0$). In Fig.~\ref{Fig4}(a), we plot $E_N$ versus the time-independent driving amplitude $E$ and atom-cavity coupling strength $G_0$. One can note that the atom-mirror entanglement in the steady state is relative small (one can also see Ref.~\cite{2008PRA77050307}) and is generated with relatively extreme system parameter (small cavity decay rate $\kappa=0.2\omega_m$). In Fig.~\ref{Fig4}(b), we present $E_N$ versus the atom-cavity coupling strength $G_0$ and mean thermal phonon occupation number $n_{\mathrm{th}}$, which shows that $E_N$ is very susceptible to thermal fluctuations of the mechanical bath and the atom-mirror entanglement can only exist in the case of low mean thermal occupation number.

\begin{figure}[H]
\centering
\includegraphics[scale=0.5]{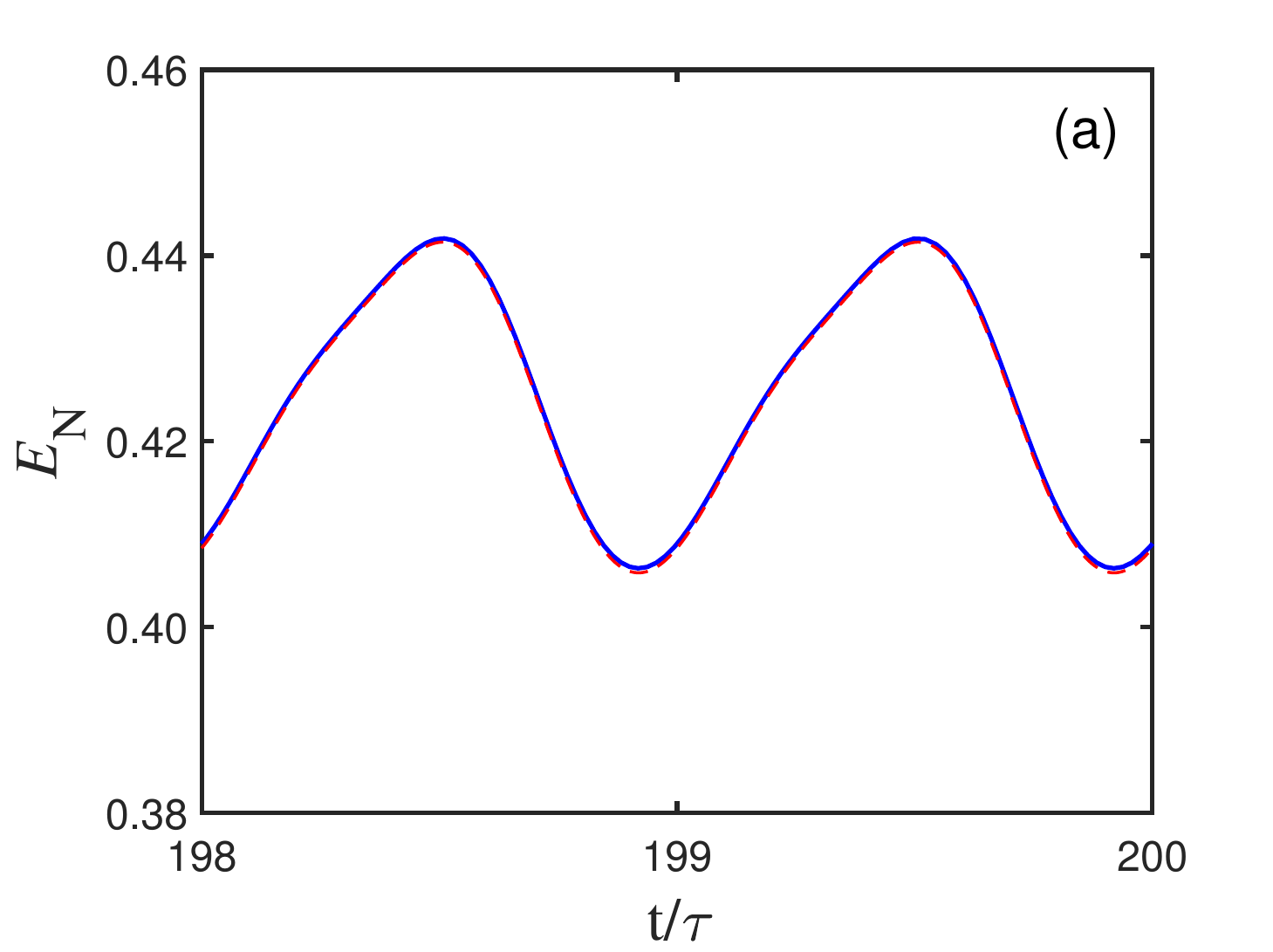}
\hspace{0.3in}
\includegraphics[scale=0.5]{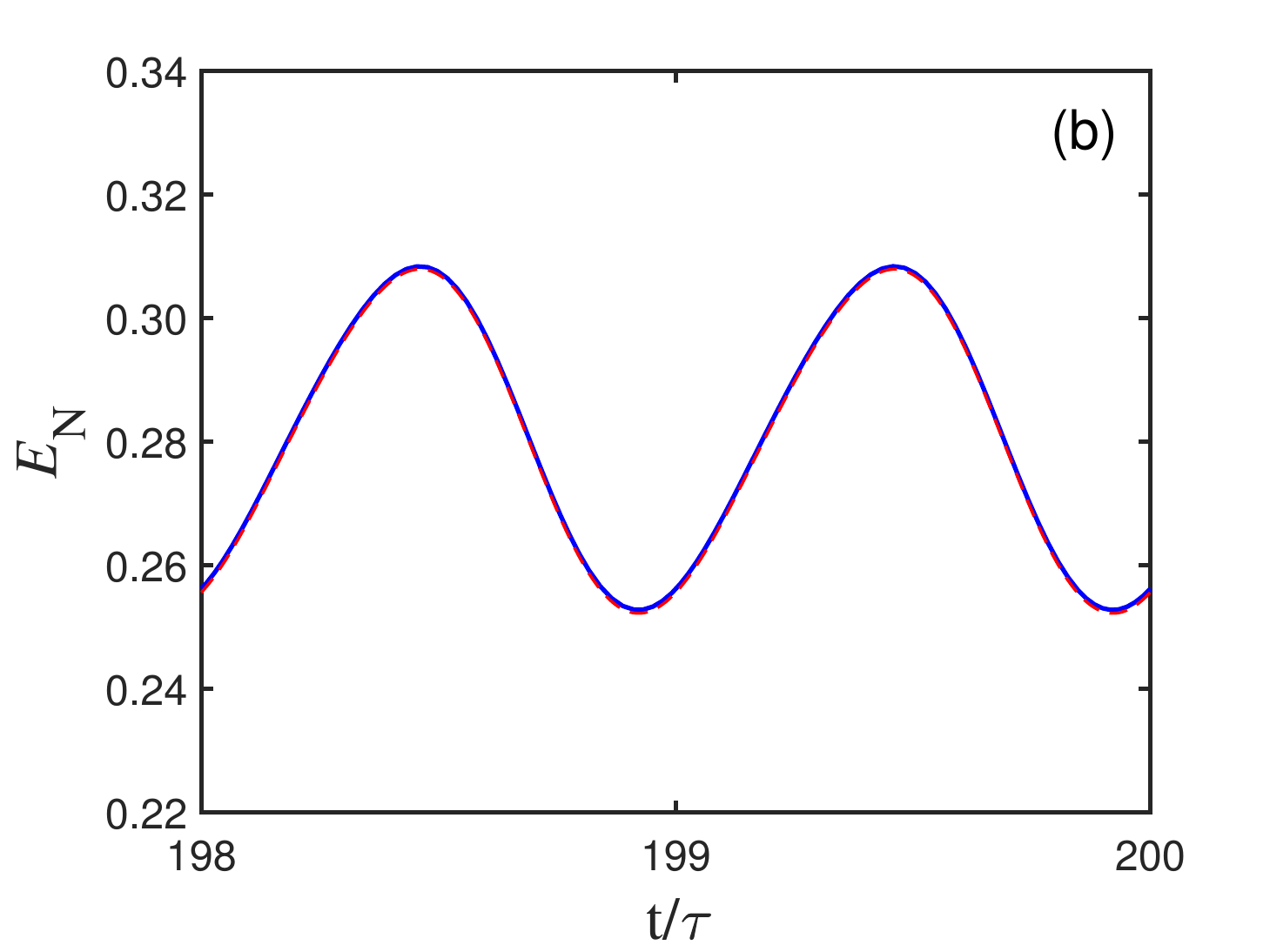}
\caption{(Color online) Atom-mirror entanglement $E_N$ with the specific modulation form of external driving as a function of time from $t=198\tau$ to $t=200\tau$ with (a) $n_{\mathrm{th}}=0$ and (b) $n_{\mathrm{th}}=50$. In both figures, the blue solid and red dashed lines correspond to the exact numerical results of first moments in Eq.~(\ref{E09}) and the analytical solutions in Eqs.~(\ref{E14}) and (\ref{E15}), respectively. All other parameters are the same as those in Fig.~\ref{Fig2}.}\label{Fig5}
\end{figure}

However, if applying the specific modulation form of external driving ($E(t)=E_{-1}e^{i\Omega t}+E_0+E_1e^{-i\Omega t}$) to drive the system, from Fig.~\ref{Fig5}(a), we note that, after long enough modulation time, $E_N$ not only obtains the same periodicity of the performed periodic modulation, but also is greatly enhanced with a more loose cavity decay rate ($\kappa=2\omega_m$). Figure \ref{Fig5}(b) shows the asymptotic time evolution of $E_N$ in the presence of thermal phonon number $n_{\mathrm{th}}=50$ where the atom-mirror entanglement almost vanishes in the case of without performing periodic modulation. As clearly illustrated in Fig.~\ref{Fig5}(b), we find that, compared to the case of without modulation, the entanglement is also more robust against the thermal fluctuations of the mechanical bath. In addition, in both Figs.~\ref{Fig5}(a) and \ref{Fig5}(b), the entanglement behaviors corresponding to the analytical solutions of first moments in Eq.~(\ref{E09}) agree with the exact results obtained from the numerical solutions of first moments in Eqs.~(\ref{E14}) and (\ref{E15}) very well.

\subsection{Atom-mirror entanglement enhancement for the specific modulation form of effective optomechanical coupling}
In above subsection, we investigate the atom-mirror entanglement enhancement via the specific modulation form of the external driving. In this subsection, we focus on the atom-mirror entanglement enhancement for the specific modulation form of effective optomechanical coupling and further to determine the explicit form of the external driving. To this end, we first assume a simple structure for the asymptotic effective time-dependent optomechanical coupling
\begin{eqnarray}\label{E29}
G(t)=G_1+G_2e^{-i\Omega t},
\end{eqnarray}
where $G_1$ and $G_2$ are time-independent positive reals and related to the external driving modulation components $E_n$. According to Eq.~(\ref{E21}), once the effective optomechanical coupling $G(t)$ is left specified, we can obtain the first moment of cavity mode $\langle a(t)\rangle=G(t)/(\sqrt{2}g)=(G_1+G_2e^{-i\Omega t})/(\sqrt{2}g)$. Then the corresponding other first moments of system and the explicit external driving can be analytically derived from Eq.~(\ref{E09}) via Laplace transform
\begin{eqnarray}\label{E30}
\langle p(t)\rangle&=&k_1e^{s_1t}+k_2e^{s_2t}+k_3e^{s_3t}+k_4e^{s_4t}, \cr\cr
\langle q(t)\rangle&=&\frac{1}{\omega_m}[-\langle\dot{p}(t)\rangle-\gamma_m\langle p(t)\rangle+g|\langle a(t)\rangle|^2], \cr\cr
\langle c(t)\rangle&=&k_5e^{s_5t}+k_6e^{s_6t}+k_7e^{s_7t}, \cr\cr
E(t)&=&\langle\dot{a}(t)\rangle+(\kappa+i\delta_a)\langle a(t)\rangle-ig\langle a(t)\rangle\langle q(t)\rangle+iG_0\langle c(t)\rangle,
\end{eqnarray}
where
\begin{eqnarray}\label{E31}
s_1&=&\frac{-\gamma_m+\sqrt{\gamma_m^2-4\omega_m^2}}{2},~~~~~s_2=\frac{-\gamma_m-\sqrt{\gamma_m^2-4\omega_m^2}}{2}, \cr\cr
s_3&=&-i\Omega,~~~~~s_4=i\Omega,~~~~~s_5=0,~~~~~s_6=s_3,~~~~~s_7=-(\gamma_a+i\Delta_c), \cr\cr
k_1&=&\frac{(G_1+G_2)^2s_1^2+(G_1^2+G_2^2)\Omega^2}{2g(s_1-s_2)(s_1-s_3)(s_1-s_4)},~~~~~
k_2=\frac{(G_1+G_2)^2s_2^2+(G_1^2+G_2^2)\Omega^2}{2g(s_2-s_1)(s_2-s_3)(s_2-s_4)}, \cr\cr
k_3&=&\frac{(G_1+G_2)^2s_3^2+(G_1^2+G_2^2)\Omega^2}{2g(s_3-s_1)(s_3-s_2)(s_3-s_4)},~~~~~
k_4=\frac{(G_1+G_2)^2s_4^2+(G_1^2+G_2^2)\Omega^2}{2g(s_4-s_1)(s_4-s_2)(s_4-s_3)}, \cr\cr
k_5&=&\frac{-iG_0(G_1+G_2)s_5+G_0G_1\Omega}{\sqrt{2}g(s_5-s_6)(s_5-s_7)},~~~~~~
k_6=\frac{-iG_0(G_1+G_2)s_6+G_0G_1\Omega}{\sqrt{2}g(s_6-s_5)(s_6-s_7)}, \cr\cr
k_7&=&\frac{-iG_0(G_1+G_2)s_7+G_0G_1\Omega}{\sqrt{2}g(s_7-s_5)(s_7-s_6)}.
\end{eqnarray}

\begin{figure}
\centering
\includegraphics[scale=0.5]{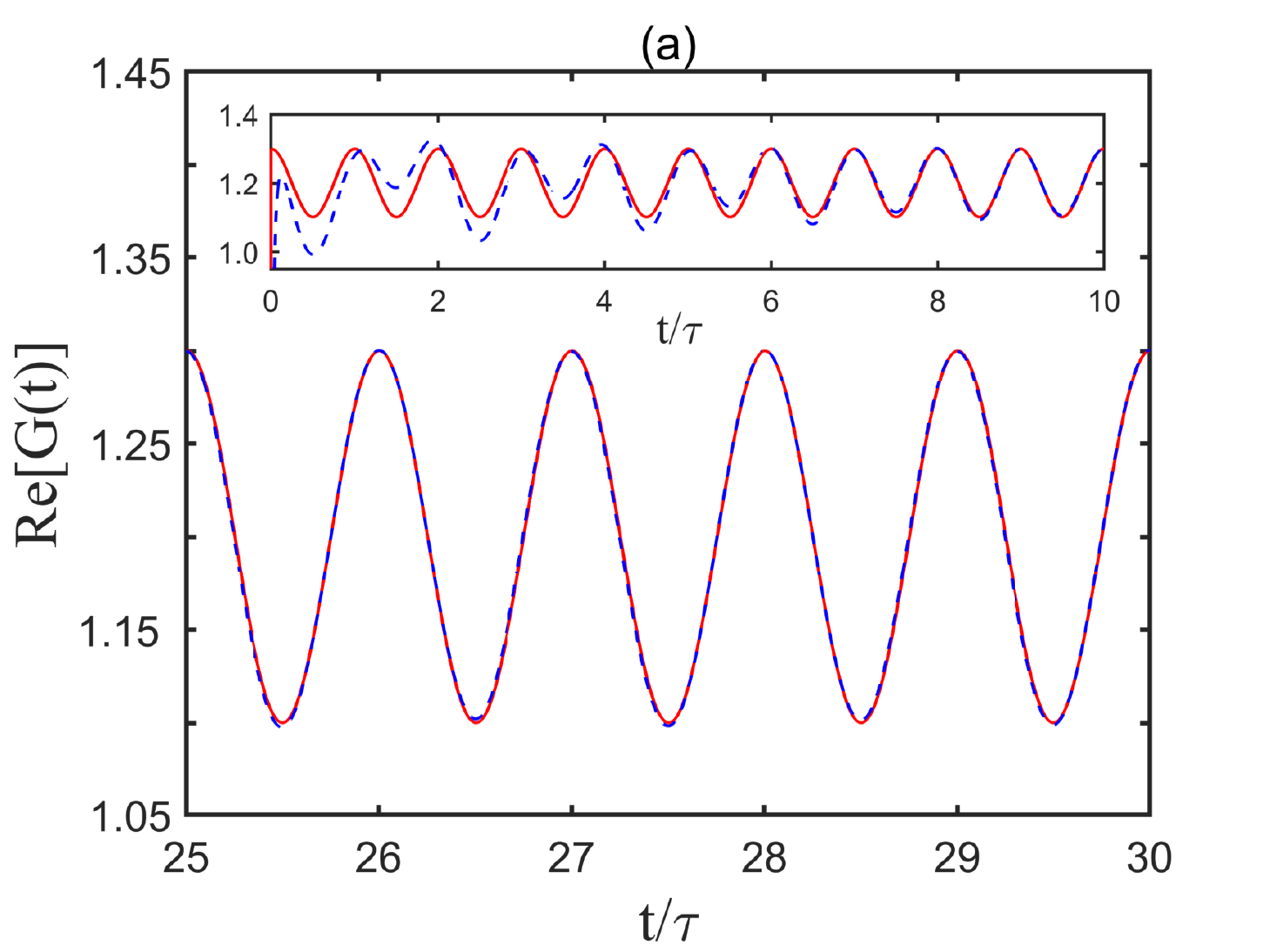}
\hspace{0.3in}
\includegraphics[scale=0.5]{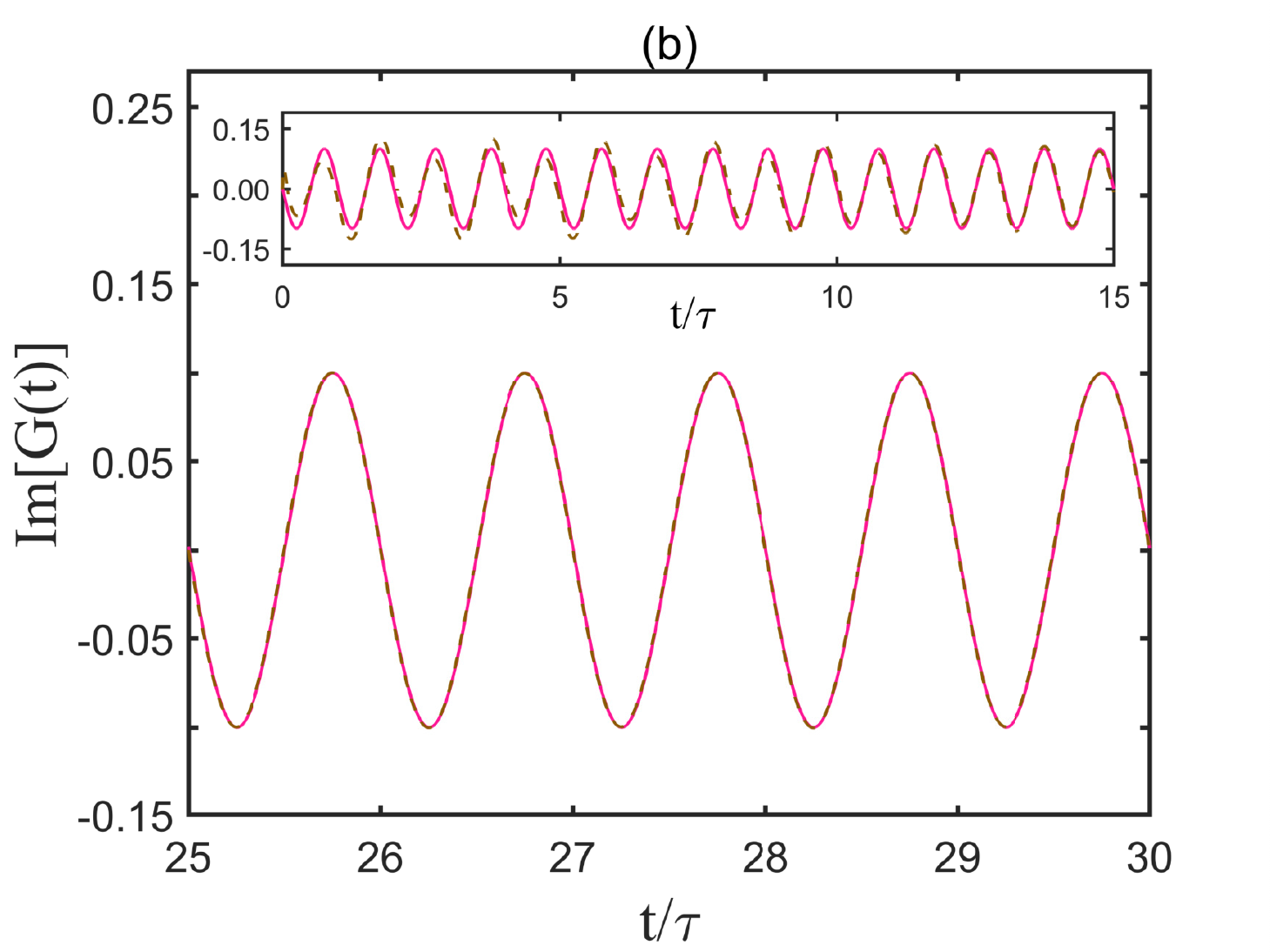}
\caption{(Color online) Asymptotic time evolution of the real and imaginary parts of effective optomechanical coupling $G(t)$. In both figures, the dashed lines represent the numerical results obtained from Eqs.~(\ref{E09}) and (\ref{E21}) by performing the particular form of external driving modulation $E(t)$ shown in Eq.~(\ref{E33}) while the solid lines refer to the specified expression $G(t)$ in Eq.~(\ref{E29}). The chosen system parameters are (in units of $\omega_m$): $\delta_a=1$, $\kappa=10$, $\gamma_m=10^{-3}$, $g=10^{-3}$, $\Delta_c=-1$, $\gamma_a=10^{-3}$, $G_0=1$, $G_1=1.2$, $G_2=0.1$, and $\Omega=2$.}\label{Fig6}
\end{figure}

In the long time limit, due to $\omega_m\gg\gamma_m(\gamma_a)>0$, one has $e^{s_1t}$, $e^{s_2t}$, and $e^{s_7t}\rightarrow0$, and the following approximations
\begin{eqnarray}\label{E32}
\langle q(t)\rangle&\simeq&\frac{1}{\omega_m}[-\langle\dot{p}(t)\rangle+g|\langle a(t)\rangle|^2], \cr\cr
k_3&\simeq&\frac{-iG_1G_2\Omega}{2g}\times\frac{1}{(-i\Omega-i\omega_m)(-i\Omega+i\omega_m)}=\frac{iG_1G_2\Omega}{2g(\Omega^2-\omega_m^2)}, \cr\cr
k_4&\simeq&\frac{iG_1G_2\Omega}{2g}\times\frac{1}{(i\Omega-i\omega_m)(i\Omega+i\omega_m)}=-\frac{iG_1G_2\Omega}{2g(\Omega^2-\omega_m^2)}.
\end{eqnarray}
Thus, we have
\begin{eqnarray}\label{E33}
\langle a(t)\rangle&=&\frac{1}{\sqrt{2}g}(G_1+G_2e^{-i\Omega t}), \cr\cr
\langle p(t)\rangle&\simeq&\frac{iG_1G_2\Omega}{2g(\Omega^2-\omega_m^2)}(e^{-i\Omega t}-e^{i\Omega t}), \cr\cr
\langle q(t)\rangle&\simeq&\frac{G_1^2+G_2^2}{2g\omega_m}-\frac{G_1G_2\omega_m}{2g(\Omega^2-\omega_m^2)}(e^{-i\Omega t}+e^{i\Omega t}), \cr\cr
\langle c(t)\rangle&\simeq&-\frac{iG_0G_1}{\sqrt{2}g(\gamma_a+i\Delta_c)}+\frac{G_0G_2}{\sqrt{2}ig[\gamma_a+i(\Delta_c-\Omega)]}e^{-i\Omega t},\cr\cr
E(t)&\simeq&E_2e^{-2i\Omega t}+E_1e^{-i\Omega t}+E_0+E_{-1}e^{i\Omega t},
\end{eqnarray}
where  the external driving modulation components $E_n$ are
\begin{eqnarray}\label{E34}
E_2&=&\frac{iG_1G_2^2\omega_m}{2\sqrt{2}g(\Omega^2-\omega_m^2)}, ~~~~~~~~~~~~~~~
E_{-1}=\frac{iG_1^2G_2\omega_m}{2\sqrt{2}g(\Omega^2-\omega_m^2)}, \cr\cr
E_1&=&\frac{G_2}{\sqrt{2}g}[\kappa+i(\delta_a-\Omega)]-\frac{iG_2}{2\sqrt{2}g\omega_m}\left(2G_1^2+G_2^2-\frac{G_1^2\Omega^2}{\Omega^2-\omega_m^2}\right)+\frac{G_0^2G_2}{\sqrt{2}g[\gamma_a+i(\Delta_c-\Omega)]}, \cr\cr
E_0&=&\frac{G_1}{\sqrt{2}g}(\kappa+i\delta_a)-\frac{iG_1}{2\sqrt{2}g\omega_m}\left(G_1^2+2G_2^2-\frac{G_2^2\Omega^2}{\Omega^2-\omega_m^2}\right)+\frac{G_0^2G_1}{\sqrt{2}g(\gamma_a+i\Delta_c)}.
\end{eqnarray}

To verify the validity of above approximations, as shown in Fig.~\ref{Fig6}, we plot the asymptotic time evolution of the real and imaginary parts of effective optomechanical coupling $G(t)$ obtained from, respectively, the numerical result by performing the particular form of external driving modulation $E(t)$ shown in Eq.~(\ref{E33}) and the specified $G(t)$ in Eq.~(\ref{E29}). The insets of Fig.~\ref{Fig6} clearly present the slow approaching process of these two results and after about 30 modulation periods, the desired periodically modulated effective optomechanical coupling $G(t)$ in Eq.~(\ref{E29}) can be indeed precisely engineered by implementing the external driving modulation components $E_n$ shown in Eq.~(\ref{E34}).

\begin{figure}
\centering
\includegraphics[width=3.0in,height=2.5in]{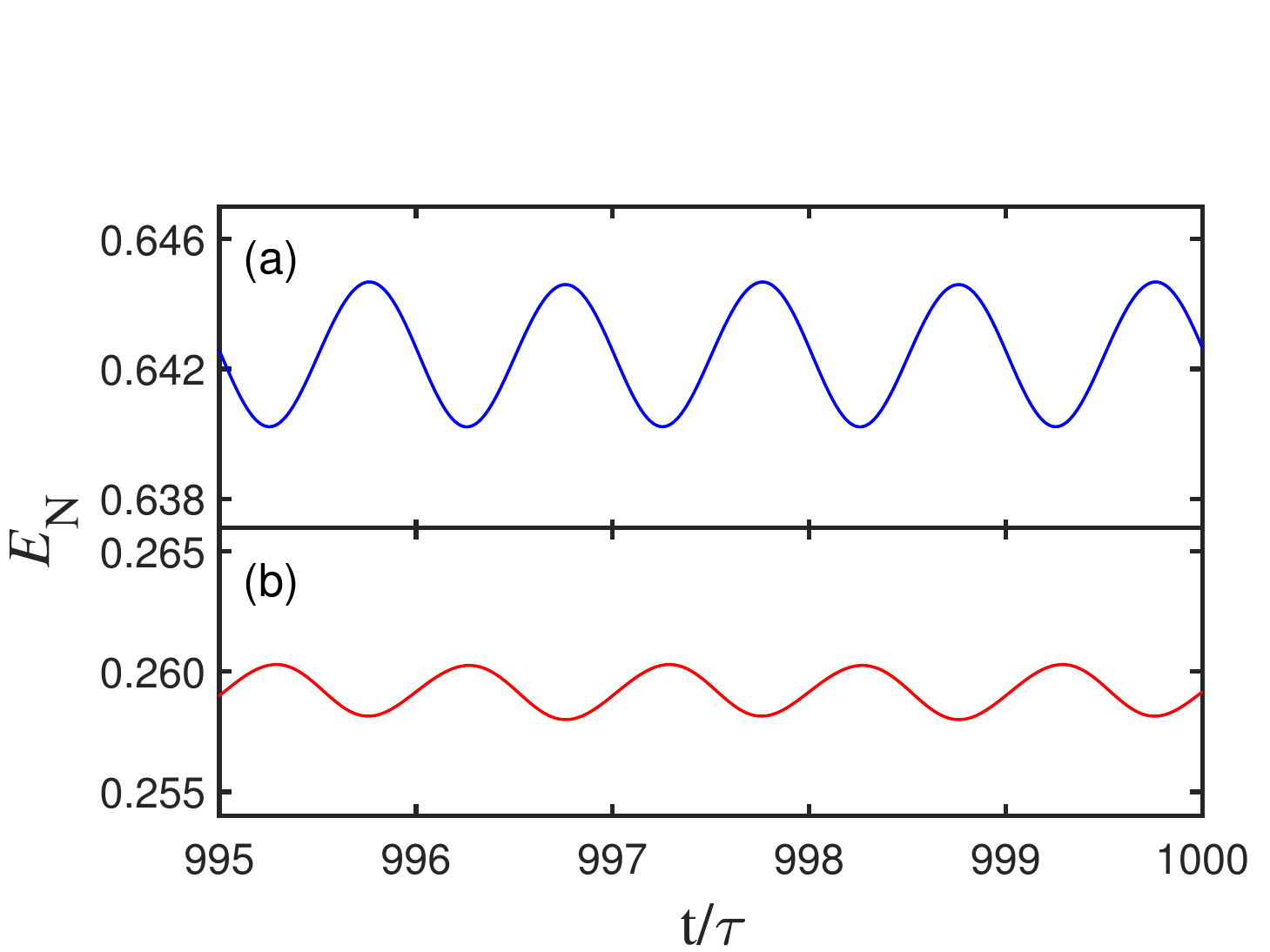}
\hspace{0.2in}
\includegraphics[width=3.0in,height=2.6in]{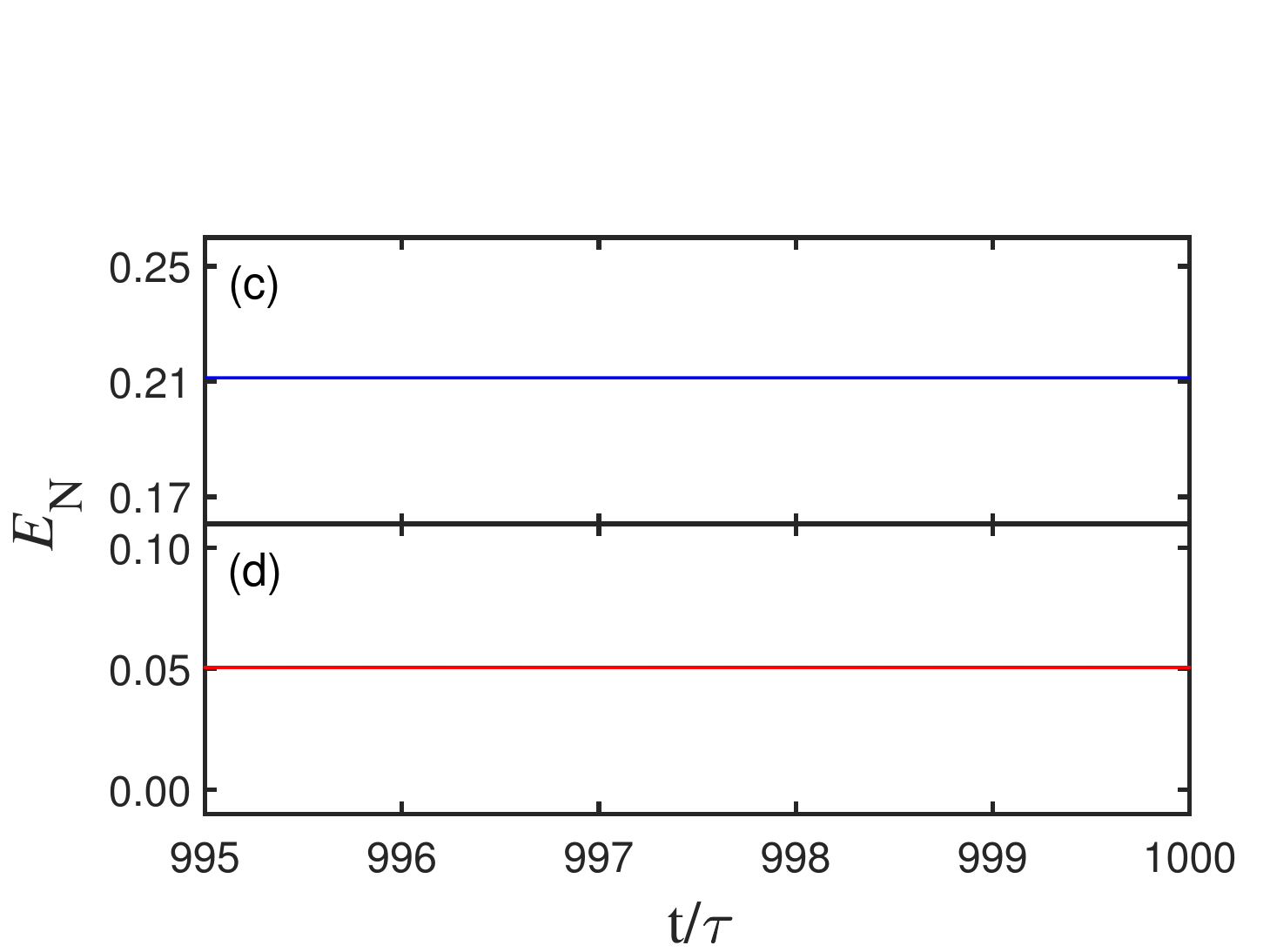}
\caption{(Color online) Atom-mirror entanglement $E_N$ in terms of periodic structure of effective optomechanical coupling as a function of time in the long time limit from $t=995\tau$ to $t=1000\tau$ with (a) $n_{\mathrm{th}}=0$ and (b) $n_{\mathrm{th}}=50$, where $E_N$ is numerically solved via the exact first moments in Eq.~(\ref{E30}). Steady-state atom-mirror entanglement $E_N$ in terms of constant structure of effective optomechanical coupling $(G_2=0)$ with (c) $n_{\mathrm{th}}=0$ and (d) $n_{\mathrm{th}}=50$.  The chosen parameters in (a) and (b) are the same as those in Fig.~\ref{Fig6} while the parameters in (c) and (d) are the same as those in Fig.~\ref{Fig6} except  $\Delta_a=1$, $\kappa=0.8$,  and $G_2=0$ (in units of $\omega_m$).}\label{Fig7}
\end{figure}

Figure \ref{Fig7} shows the atom-mirror entanglement behaviors in the cases of the periodic and constant structures as for the effective optomechanical coupling. From Fig.~\ref{Fig7}, one can find that, in the constant structure, the atom-mirror entanglement is weak and can only exist in the low mean thermal phonon number limit with the relatively small cavity decay rate yet ($\kappa=0.8\omega_m$). Instead, once applying the periodic structure, the entanglement is not only significantly enhanced with a more loose cavity decay rate ($\kappa=10\omega_m$), but also is more resistant to thermal fluctuations of the bath.

\section{Mechanical squeezing generation in the unresolved sideband regime via modulation}\label{Sec6}

\begin{figure}
\centering
\includegraphics[scale=0.4]{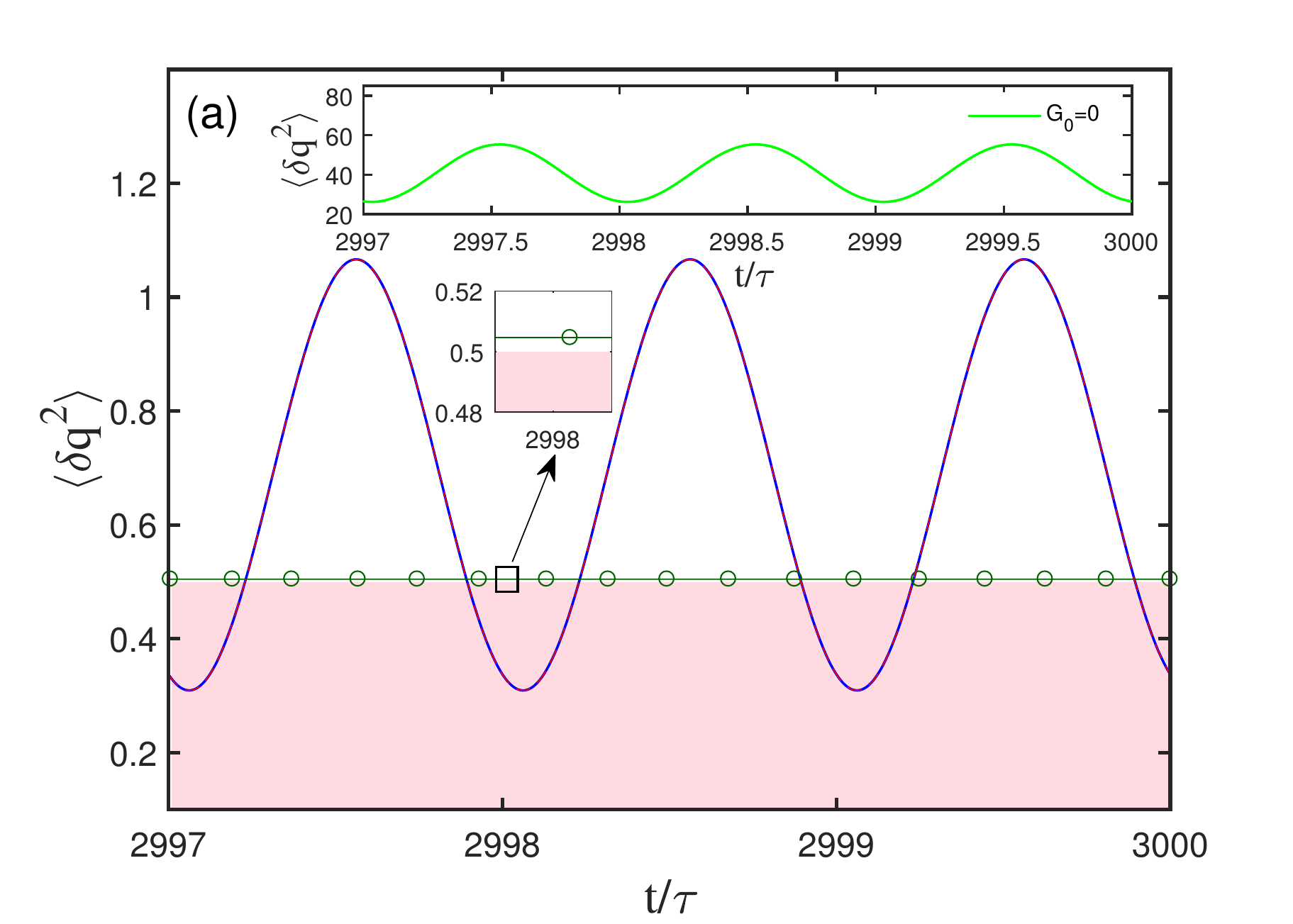}
\hspace{0.3in}
\includegraphics[scale=0.4]{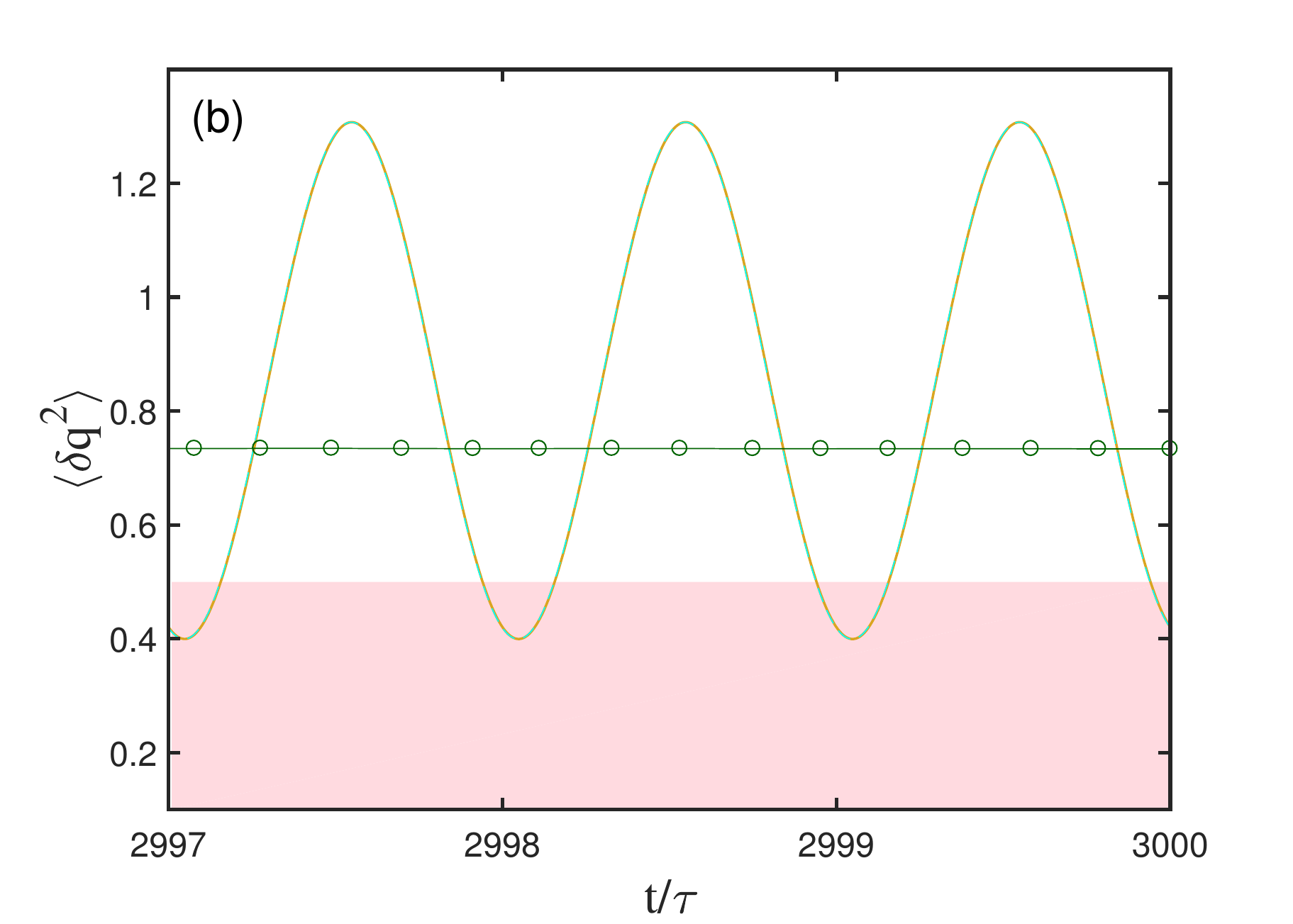}
\caption{(Color online) Variance of the mechanical position operator with (a) $n_{\mathrm{th}}=0$ and (b) $n_{\mathrm{th}}=100$. In both figures, the circle line refers to the result in the absence of modulation ($\Omega=0$), and the solid and dashed lines correspond to the exact numerical results of first moments in Eq.~(\ref{E09}) and the analytical solutions in Eqs.~(\ref{E14}) and (\ref{E15}), respectively. The inset in (a) presents  variance of the mechanical position operator corresponding to a pure optomechanical system (there isn't atomic ensemble in system). The pink area represents the region below quantum noise limit. The chosen system parameters are (in units of $\omega_m$): $\delta_a=1$, $\kappa=10$, $\gamma_m=10^{-6}$, $g=5\times10^{-5}$, $\Delta_c=-1.1$, $\gamma_a=10^{-3}$, and $G_0=6$. The lowest external driving modulation components $E_0=12\times10^4$ and $E_{-1}=E_1=2\times10^4$, and modulation frequency $\Omega=2$.}\label{Fig8}
\end{figure}

We now turn to investigate the mechanical squeezing generation in the unresolved sideband regime ($\kappa>\omega_m$) via periodic modulation. The variances of the quantum fluctuations around the first moments for the mechanical position and momentum operators are in the form $\langle\delta q^2\rangle-\langle\delta q\rangle^2$ and $\langle\delta p^2\rangle-\langle\delta p\rangle^2$, respectively. The mechanical oscillator is squeezed if either $\langle\delta q^2\rangle-\langle\delta q\rangle^2$ or $\langle\delta p^2\rangle-\langle\delta p\rangle^2$ is less than 1/2. Due to $\langle\delta q\rangle=0$ and $\langle\delta p\rangle=0$, so the equation of motion for the CM in Eq.~(\ref{E23}) can also completely characterize the time evolution of variances for the mechanical position and momentum operators ($\langle\delta q^2\rangle$ ($\langle\delta p^2\rangle$) just is the matrix element of CM $V_{11}$ ($V_{22}$)).

\begin{figure}
\centering
\includegraphics[scale=0.6]{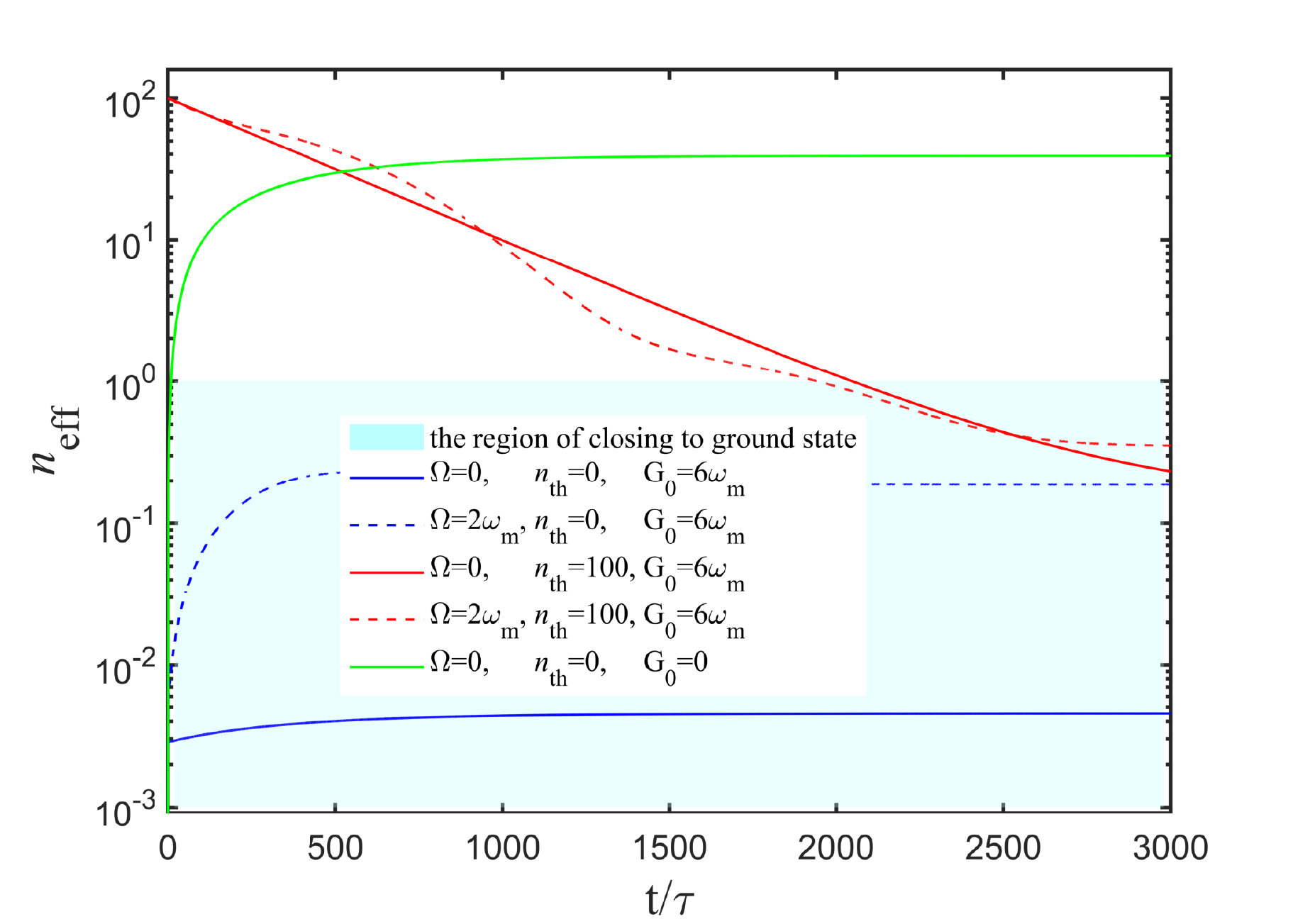}
\caption{(Color online) The time evolution of the mean phonon number $n_{\mathrm{eff}}$. For the system parameters, see the Fig.~\ref{Fig8}.}\label{Fig9}
\end{figure}

In Fig.~\ref{Fig8}, we plot the time evolution of variance for the mechanical oscillator position operator from $t=2997\tau$ to $t=3000\tau$ in the periodic driving modulation regime. As a comparison, we first present the result in the case of without periodic modulation ($\Omega=0$). For $\Omega=0$, the system turns into constant driving regime and when the initial mean thermal phonon number $n_{\mathrm{th}}=0$, as the blue solid line depicted in Fig.~\ref{Fig9}, due to the quantum interference mechanism induced by atoms inside the cavity~\cite{2015PRA92033841,2009PRA80061803,2015JOSAB322314,2018OE266143}, the mechanical oscillator can always be retained close to ground state in the total evolution precess even though $\kappa>\omega_m$. Even if the initial mean thermal phonon number $n_{\mathrm{th}}=100$, as shown by the red solid line, the mechanical oscillator can still be cooled close to ground state with the time evolution finally. Since the mechanical oscillator is prepared in an approximate vacuum state at the time interval $[2997\tau, 3000\tau]$,  as the circle lines shown in Fig.~\ref{Fig8}, the variance of the mechanical position operator without modulation is near to the quantum noise limit. However, for $\Omega=2\omega_m$, when we appropriately modulate the external driving components $E_n$, the mechanical position is periodically squeezed over time with the same period of modulation. In addition, from the dashed lines in Fig.~\ref{Fig9}, we find that the mechanical oscillator can also be cooled close to its ground state, thus the mechanical oscillator will be in a squeezed thermal state instead of a pure squeezed vacuum. Here the mechanical squeezing results from the dynamics resembles the effect of parametric amplification and the spring constant is seems modulated at twice the mechanical resonance frequency~\cite{2009Physics,2009PRL103213603,2012PRA86013820,2013PRA88063833}.

As is well known, in a standard optomechanical system, it is hard to realize ground-state cooling without the help of other auxiliary systems or manipulation means. It is obvious that the green line in Fig.~\ref{Fig9} also verifies this point. If we withdraw the atoms from optomechanical cavity, even though the initial mean thermal phonon number $n_{\mathrm{th}}=0$, the mean phonon number $n_{\mathrm{eff}}$ rapidly increases up to about 40 and the mechanical oscillator finally reaches the dynamical equilibrium with the bath. This is because the quantum interference mechanism caused by atoms is broken and the thermal noise of mechanical bath significantly affects the finial phonon population of mechanical oscillator. As shown in the inset of Fig.~\ref{Fig8}(a), it can be predicted that, even though there exists the periodic driving modulation and the initial mean thermal phonon number $n_{\mathrm{th}}=0$, the variance of the mechanical position operator has been far away from the quantum noise limit. Hence, the thermal noise of mechanical bath unavoidably suppresses the emergence of quantum effects.

\begin{figure}[H]
\centering
\includegraphics[scale=0.9]{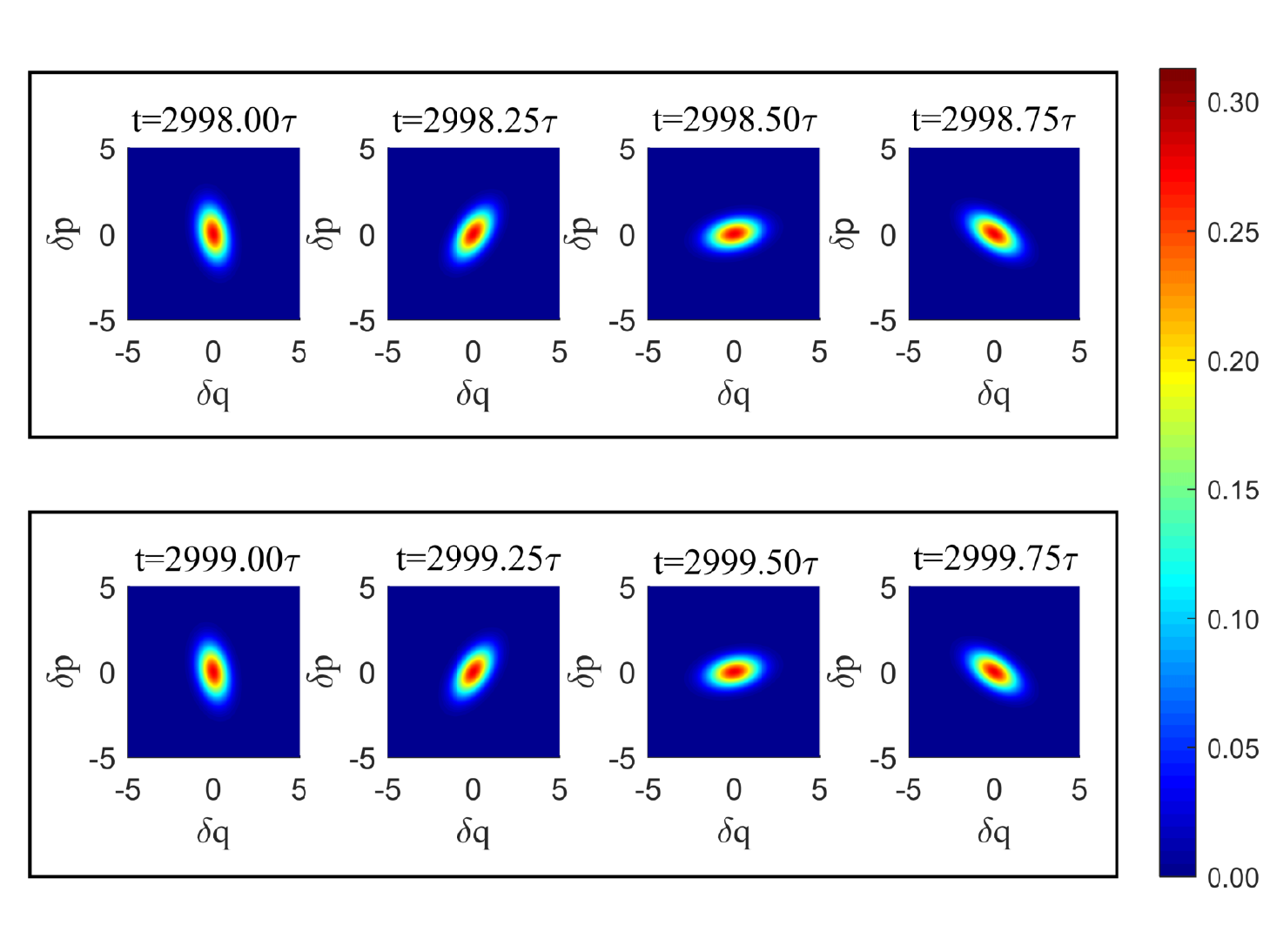}
\caption{(Color online) The Wigner function for the mechanical mode at some different specific times. The parameters are the same as those in Fig.~\ref{Fig8}(a).}\label{Fig10}
\end{figure}

\begin{figure}[H]
\centering
\includegraphics[scale=0.6]{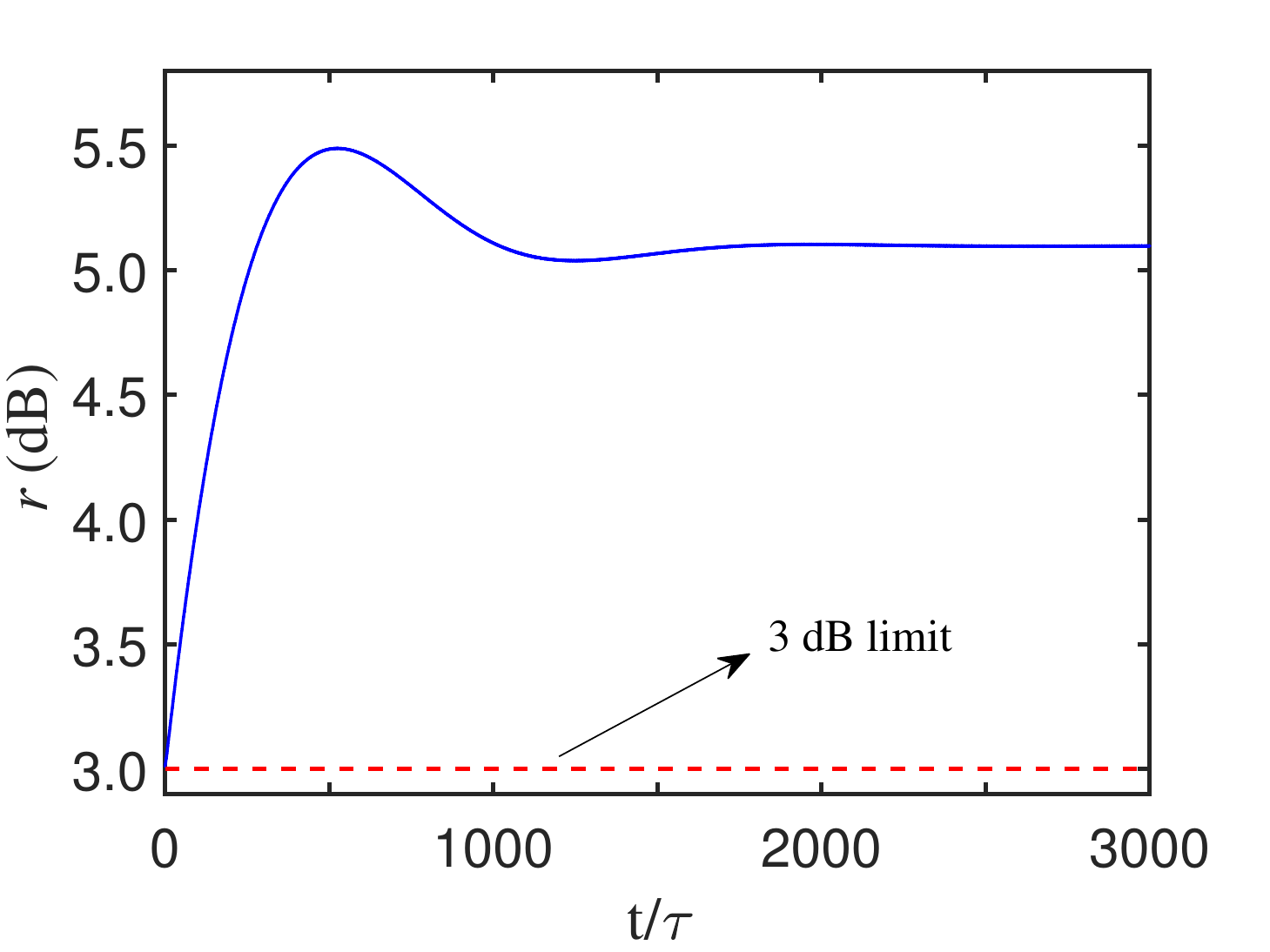}
\caption{(Color online) The time evolution of the single mode squeezing parameter $r$. The parameters are the same as those in Fig.~\ref{Fig8}(a).}\label{Fig11}
\end{figure}

In order to further illustrate the features of the mechanical squeezing, as shown in Figs.~\ref{Fig10} and \ref{Fig11}, we depict the respective Wigner functions (see Appendix~\ref{App2}) at some different specific times in the phase space and the time evolution of the single mode squeezing parameter $r$ (see Appendix~\ref{App3}), respectively. Since the squeezing parameter $r$ is a constant in the long time limit, one can observe from Fig.~\ref{Fig10} that the shape of the respective Wigner functions is fixed, which means that the mechanical oscillator is always squeezed but, due to the periodic modulation, the direction of squeezing continuously rotates in phase space with the same period of modulation.

\section{Conclusions}\label{Sec7}
In conclusion, we have investigated the quantum effects of atom-mirror entanglement enhancement and mechanical squeezing induced by the periodically modulated driving in a hybrid optomechanical system consisting of the atomic ensemble and a standard optomechanical cavity. It has been shown that the system will obtain the same period of the modulation in the long time limit. Compared to the case of no modulation, the atom-mirror entanglement generated for the specific modulation forms of external driving and effective optomechanical coupling is greatly enhanced with more loose cavity decay rate and is more robust against thermal fluctuations of mechanical bath. The desired form of periodically modulated effective optomechanical coupling can also be precisely engineered by the external driving modulation components. The mechanical squeezing induced by the periodically modulated driving can be generated successfully in the unresolved regime via resorting to the quantum interference mechanism caused by the atoms inside the cavity and choosing appropriately the modulated driving amplitude. From the Winger functions at some different specific times and the time evolution of squeezing parameter in the long time limit, we find that the mechanical oscillator is always squeezed but, due to the periodic modulation, the direction of squeezing rotates continuously in phase space with the same period of modulation. The present scheme may benefit forward preparing significantly entangled macroscopic objects with available techniques in present cavity optomechanics and the possible  ultraprecise detection applications involving mechanical squeezing based on cavity optomechanics.

\newpage
\begin{center}
$\mathbf{Acknowledgements}$
\end{center}

This work was supported by the National Natural Science Foundation of China under Grant
Nos. 11465020, 11264042, 61465013, 61575055, and the Project of Jilin Science and Technology Development for Leading Talent of Science and Technology Innovation in Middle and
Young and Team Project under Grant No. 20160519022JH.

\appendix
\section{Derivation of the equation of motion for the CM}\label{App1}
The formal solution of Eq.~(\ref{E18}) is
\begin{eqnarray}\label{A1}
u(t)=U(t, t_0)u(t_0)+\int_{t_0}^tU(t, s)n(s)ds,
\end{eqnarray}
where $U(t, t_0)$ is the principal matrix solution of the homogeneous system which satisfies $\dot{U}(t, t_0)=A(t)U(t, t_0)$ and $U(t_0, t_0)=\mathds{1}$. The differential of matrix element for the CM is
\begin{eqnarray}\label{A2}
\frac{dV_{k,l}}{dt}&=&\left\langle\frac{du_k}{dt}u_l+u_k\frac{du_l}{dt}+\frac{du_l}{dt}u_k+u_l\frac{du_k}{dt}\right\rangle/2 \cr\cr
&=&\Bigg\langle\left[\sum_nA_{k,n}(t)u_n(t)+n_k(t)\right]u_l(t)+u_k(t)\left[\sum_nA_{l,n}(t)u_n(t)+n_l(t)\right] \cr\cr
&&+\left[\sum_nA_{l,n}(t)u_n(t)+n_l(t)\right]u_k(t)+u_l(t)\left[\sum_nA_{k,n}(t)u_n(t)+n_k(t)\right]\Bigg\rangle/2 \cr\cr
&=&\sum_nA_{k,n}(t)\langle u_n(t)u_l(t)+u_l(t)u_n(t)\rangle/2+\sum_nA_{l,n}(t)\langle u_k(t)u_n(t)+u_n(t)u_k(t)\rangle/2 \cr\cr
&&+\langle n_k(t)u_l(t)+u_k(t)n_l(t)+n_l(t)u_k(t)+u_l(t)n_k(t)\rangle/2 \cr\cr
&=&\sum_nA_{k,n}(t)V_{n,l}(t)+\sum_nA_{l,n}(t)V_{k,n}(t) \cr\cr
&&+\langle n_k(t)u_l(t)+u_k(t)n_l(t)+n_l(t)u_k(t)+u_l(t)n_k(t)\rangle/2.
\end{eqnarray}
Substituting the matrix elements of Eq.~(\ref{A1})
\begin{eqnarray}\label{A3}
u_l(t)=\sum_n\Big[U_{l,n}(t, t_0)u_n(t_0)+\int_{t_0}^tU_{l,n}(t,s)n_n(s)ds\Big], \cr\cr
u_k(t)=\sum_n\Big[U_{k,n}(t, t_0)u_n(t_0)+\int_{t_0}^tU_{k,n}(t,s)n_n(s)ds\Big], 
\end{eqnarray}
into Eq.~(\ref{A2}) and utilizing the noise correlation relations in Eqs.~(\ref{E06}) and (\ref{E08}), one can obtain
\begin{eqnarray}\label{A4}
\frac{dV_{k,l}}{dt}=\sum_nA_{k,n}(t)V_{n,l}(t)+\sum_nA_{l,n}(t)V_{k,n}(t)+D_{k,l},
\end{eqnarray}
where $D=\mathrm{diag}[0, \gamma_a(2n_{\mathrm{th}}+1), \kappa, \kappa, \gamma_a, \gamma_a]$. The above equation is written in terms of matrix:
\begin{eqnarray}\label{A5}
\frac{dV}{dt}=A(t)V(t)+V(t)A^T(t)+D,
\end{eqnarray}
Eq.~(\ref{A5}) just is the equation of motion for the CM which completely describes the dynamics of the quantum fluctuations.

\section{The Wigner function of Gaussian states}\label{App2}
In terms of quantum information with continuous-variable systems, Gaussian states play an important role. They have Gaussian Wigner characteristic function and are fully characterized by the first and second moments of the quadrature operators. For a Gaussian state of $N$ bosonic modes, the first moment is denoted by the vector of mean values
\begin{eqnarray}\label{B1}
\langle R\rangle=\mathrm{Tr}(\varrho R)=[\langle R_1\rangle, \langle R_2\rangle, \cdots, \langle R_N\rangle]^T,
\end{eqnarray}
while the second moment is denoted by the $2N\times2N$ CM $V$ with element
\begin{eqnarray}\label{B2}
V_{i,j}=\frac12\langle R_iR_j+R_jR_i\rangle-\langle R_iR_j\rangle,
\end{eqnarray}
in which $R=[q_1, p_1, \cdots, q_N, p_N]^T$ is the vector of the quadrature operators. 

In the main text, we have set the first moments to zero for simplicity because following a local unitary transformation cannot affect any information-related properties. In this case, the Wigner function for a $N$ mode Gaussian state can be readily written as~\cite{2005RMP77513}
\begin{eqnarray}\label{B3}
W(R)=\frac{1}{(2\pi)^N\sqrt{\mathrm{Det}V}}\mathrm{exp}\left\{-\frac12R^TV^{-1}R\right\}.
\end{eqnarray}
\section{The single mode squeezing parameter}\label{App3}
If the CM of the bosonic mode is written as in the following form
\begin{eqnarray}\label{C1}
\sigma=
\begin{bmatrix}
\sigma_{11}~ & ~\sigma_{12} \\
\sigma_{21}~ & ~\sigma_{22}
\end{bmatrix},
\end{eqnarray}
and $\lambda$ is the minimum eigenvalue of the CM
\begin{eqnarray}\label{C2}
\lambda=\mathrm{min}\{\lambda_1, \lambda_2\},
\end{eqnarray}
the single mode squeezing parameter $r$ can be measured in terms of $\lambda$ effectively:
\begin{eqnarray}\label{C3}
r=-10\log_{10}\lambda.
\end{eqnarray}

\end{document}